\documentclass[11pt]{article}

\usepackage[a4paper,margin=0.8in]{geometry}
\usepackage{authblk}

\usepackage{xcolor}
\usepackage{import}
\usepackage{graphicx}
\usepackage{float}
\usepackage[nocompress]{cite}
\usepackage{mathtools}
\usepackage{amsmath,amsfonts,amssymb,amsthm}

\theoremstyle{plain}
\newtheorem{thm}{Theorem}[section]
\newtheorem{lem}[thm]{Lemma}
\newtheorem{prop}[thm]{Proposition}
\newtheorem{assum}[thm]{Assumption}
\theoremstyle{definition}
\newtheorem{defn}[thm]{Definition}
\newtheorem{exmp}[thm]{Example}
\theoremstyle{remark}
\newtheorem{rem}[thm]{Remark}
\newenvironment{pf*}[1]{\begin{proof}[#1]}{\end{proof}}

\usepackage{xparse}

\newcommand{\twolineright}[1]{\mathmakebox[\displaywidth][r]{\displaystyle #1}\notag\\}
\newcommand{\twolineequationprocess}[1]{%
	\begin{gather}
		#1
		\vphantom{\displaystyle\int}\ifx\twolineequationlabel\empty\else\label{\twolineequationlabel}\fi
	\end{gather}%
}

\newcommand{\twolineequationstarprocess}[1]{%
	\renewcommand{\twolineright}[1]{\mathmakebox[\displaywidth][r]{\displaystyle ##1}\notag}%
	\begin{gather*}
		#1
	\end{gather*}%
}
\newenvironment{twolineequationstar}{%
	\csname collect@body\endcsname\twolineequationstarprocess
}{}
\expandafter\let\csname twolineequation*\endcsname\twolineequationstar
\expandafter\let\csname endtwolineequation*\endcsname\endtwolineequationstar

\begin{document}

\title{Memory-Dependent Interval Markov Chain Abstractions of Stochastic Dynamics\thanks{This paper was not presented at any IFAC meeting. Corresponding author: Menno van Zutphen. This research is part of the research program SYNERGIA (project number 17626), which is partly financed by the Dutch Research Council (NWO).}}

\author[1]{Menno van Zutphen\\\texttt{m.j.t.c.v.zutphen@tue.nl}}
\author[2]{Adrien Banse\\\texttt{abanse@control.ee.ethz.ch}}
\author[1]{Domagoj Herceg\\\texttt{d.herceg@tue.nl}}
\author[3]{Giannis Delimpaltadakis\\\texttt{ioannis.delimpaltadakis@ai4i.it}}
\author[1]{Duarte J. Antunes\\\texttt{d.antunes@tue.nl}}
\affil[1]{Control Systems Technology, Department of Mechanical Engineering, Eindhoven University of Technology, P.O. Box 513, 5600 MB Eindhoven, The Netherlands}
\affil[2]{Automatic Control Laboratory, ETH Zurich, Department of Information Technology and Electrical Engineering, Physikstrasse 3, 8092 Zurich, Switzerland}
\affil[3]{Robust and Intelligent Autonomous Systems Lab, The Italian Institute of Artificial Intelligence for Industry (AI4I), Corso Castelfidardo 22, 10129 Turin, Italy}
\date{}

\maketitle

\begin{abstract}
	Finite-state \emph{interval Markov-chain} (IMC) abstractions provide sound verification and performance bounds for continuous-state stochastic systems by enclosing cell-to-cell transition probabilities and costs in intervals. Finite-state Markovian abstractions are generally lossy, as state aggregation often destroys the Markov property. Recently, \emph{Memory-dependent} Markov-chain (MC) abstractions of stochastic systems have been developed to mitigate this by recording recently visited cells. In this paper, we develop the memory-dependent extension of IMC abstractions. Building on the insight that memory narrows the family of state distributions that could occur inside the current cell, we prove that memory tightens the local intervals, at the price of a larger abstraction. We also formulate \emph{no-memory} IMC abstractions, tightening classic IMC abstraction intervals without enlarging the state space. To characterize spatial refinement relative to memory, we derive two cost-guarantee tightness upper bounds. The bounds depend, respectively, on cell size, and on the worst-case remaining distributional ambiguity after filtering over the remembered past. We then specialize our result to linear-Gaussian systems, for which we derive a closed-form upper bound on this ambiguity. Numerical examples show that the memory-dependent and no-memory constructions can produce tighter expected-cost intervals than comparable partition refined models.
\end{abstract}


\section{Introduction} \label{sec:introduction}

In recent years, autonomous systems are being developed for increasingly unpredictable environments. Examples of systems subject to such high levels of randomness include self-driving vehicles, rescue robots, smart grids, and intelligent buildings \cite{Alamir2022,Pairet2022}. Towards safe deployment of such autonomous systems, mathematical methods are needed that can formally verify whether they satisfy prespecified requirements, such as safety or performance \cite{Knight2002,Lee2016}.

\emph{Stochastic models} are employed to model this unpredictability. However, a precise quantitative method of propagating the stochasticity over time is often unavailable for those systems as a result of, e.g., strong nonlinearities, hindering formal verification and performance analysis.
\emph{Finite-state abstractions} address this difficulty by replacing the \emph{concrete} continuous stochastic process by an \emph{abstract} discrete stochastic process, while retaining quantitative guarantees on the properties of interest (see \cite{Lavaei2022Survey} for a survey). Most methods in this literature rely on partitioning the state space, and a number of works have studied this setting through partition-dependent error bounds~\cite{Soudjani2015Quantitative}, adaptive and sequential gridding procedures~\cite{Soudjani2013Adaptive}. The fundamental literature is also supported by software implementations such as \texttt{{FAUST}$^{2}$} \cite{Soudjani2015FAUST2}.

In abstractions, as the current concrete state is not exactly known (only the region in which it lies is), the transition probabilities are themselves uncertain. \emph{Interval Markov chains} (IMCs) --- in the autonomous case --- and \emph{Interval Markov Decision Processes} (IMDPs) --- in the controlled case --- encode this uncertainty through lower and upper bounds on these transition probabilities~\cite{Givan2000,Delimpaltadakis2022IntervalMD}. This then induces robust verification and synthesis problems over the set of all transition distributions consistent with the intervals. They have been used for scalable synthesis with explicit interval uncertainty~\cite{Cauchi2019Efficiency}, $\omega$-regular objectives~\cite{Dutreix2022Omega}, abstractions based on noise partitioning~\cite{Skovbekk2023Noise}, and approximate relations between uncertain finite-state models~\cite{Lun2018Approximate,Mathiesen2024}. In this paper, we focus on finite-horizon expected cumulative cost problems.

% In this paper, the IMC viewpoint is useful as it gives direct lower and upper expected-cost guarantees through robust dynamic programming.

The precision of an interval abstraction depends on the information retained by its abstract states. Partition refinement is the standard way of increasing this information. Smaller cells reduce the variation of the concrete transition kernel inside each abstract state, tightening the corresponding probability intervals. This principle underlies adaptive gridding and related refinement methods, and also appears in other control-theoretic routes based on approximate bisimulation~\cite{Zamani2014SymbolicControl}, discretization-free representations~\cite{Zamani2015Switched}, or compositional abstraction relations for interconnected systems~\cite{Lavaei2020Compositional}. These approaches primarily improve the abstraction by changing the geometric partition, the abstraction relation, or the system decomposition.

However, state aggregation generally makes the induced process of visited cells non-Markovian; the distribution of the next cell may depend on past cells beyond the current one. \emph{Memory-dependent abstractions} remember past abstract cells rather than only considering the current cell. In the deterministic setting, such models have been studied in the context of \emph{behavioural systems} (see e.g. the so-called \emph{$\ell$-complete} abstractions in \cite{Moor1999,Moor2002,SchmuckRaisch2014,Yang2020}) and more recently in data-driven settings (see e.g. \cite{Coppola2023DataDriven,pmlr-v211-banse23a,Banse2025CantorKantorovich}). In \cite{Banse2025Transfer}, Banse et al.\ developed a similar memory-dependent viewpoint for stochastic systems, but constructing (labeled) Markov chains instead of IMCs, thereby leading to approximation errors, which are quantified there through a transfer-operator formulation. Taken together, these results extend the memory-dependent intuition to stochastic systems. Yet, the guarantees they offer rest on a-poteriori quantification of approximation error, instead of following directly from robust dynamic programming on the abstraction itself. Such error quantifications moreover rely on restrictive regularity assumptions on the system, and tend to be conservative.

The present paper studies precisely this interval-valued counterpart. To achieve this, we introduce a novel construction of IMC abstractions based on filtering ideas. Without memory, an abstract observation exclusively restricts the conditional measure of the concrete state to be concentrated inside the current cell. With memory, a sequence of abstract observations restricts the conditional measure within the current cell to a parameterized family of measures, leading to tighter costs and intervals, at the expense of increased computational complexity. The main purpose of this paper is thus to develop and analyze such memory-dependent IMC abstractions for finite-horizon expected cumulative cost problems.

Partition refinement reduces the diameter of the current-state cells, whereas memory restricts the conditional measure of the concrete state inside the current cell. This distinction enables separate analyses of their tightening mechanisms and system-dependent upper bounds without asserting a general ordering between them.

The paper makes the following contributions.
\begin{enumerate}
	\item We develop memory-dependent IMC abstractions from a filtering perspective. We show that filtering restricts the family of admissible conditional state measures, and that adding memory narrows these families, resulting in tighter stage-cost and transition intervals. Classic IMC abstractions are recovered as a special case of memory-dependent IMC abstractions with memory length zero.
	\item To characterize the effect of memory lengthening relative to partition refinement, we derive separate upper bounds on the \emph{robust value-function width}, defined as the maximum statewise gap between the lower and upper robust cost-to-go values. For a fixed horizon, the partition-refinement upper bound vanishes as the refined-cell diameter converges to zero. For linear-Gaussian dynamics, the memory-dependent bound vanishes as $\ell\to\infty$ along full-memory stages with fixed remaining horizon when the analytically computable coefficient $\kappa$ satisfies $\kappa<1$. The bounds identify two tightening mechanisms: partition refinement reduces the refined-cell diameter, whereas memory reduces the worst-case remaining distributional ambiguity induced by the remembered past.
	\item We compare the effect of memory lengthening and partition refinement through numerical experiments on a linear-Gaussian benchmark  under an equal number of stored transition probability intervals for fair comparison. We demonstrate that memory-dependent constructions can significantly outperform partition refinement on both the linear-Gaussian benchmark and a simple non-linear system. Yet, we conclude that neither of the two modalities is universally superior.
\end{enumerate}

% The organization of the paper is as follows. Section~\ref{sec:concretesystemandabstractionproblem} defines IMCs and their robust dynamic program before formulating the continuous-state abstraction problem. Section~\ref{sec:memory_dependent_imcs} develops the memory-dependent family and analyzes source-cell partition refinement alongside it. Section~\ref{sec:refinement_efficiency} derives these upper bounds separately and develops the linear-Gaussian specialization. Section~\ref{sec:linear_gaussian_benchmark} presents numerical comparisons. Section~\ref{sec:conclusion} concludes the paper and discusses directions for further work.

\section{Concrete system and abstraction problem} \label{sec:concretesystemandabstractionproblem}

In this section, we formulate the formal abstraction problem studied in the paper. We first introduce the concrete, \emph{continuous-state Markov chain} (MC). We then outline the problem of obtaining expected-cost soundness guarantees for \emph{interval Markov chain} (IMC) abstractions, which are finite-state Markov-chains with interval-valued transitions and costs, subjected to robust dynamic programming. In contrast to existing approaches in the literature, we consider memory-dependent IMC constructions, exploiting memory of the $\ell\in\mathbb{N}$ previous abstract cells to improve the tightness of the resulting bounds.

\subsection{Preliminaries}\label{sec:preliminaries}

For a measurable set $A$, let $\mathcal{P}(A)$ denote the set of probability measures on $A$. All models considered in this paper share a fixed finite horizon $H\in\mathbb{N}_{\ge1}$, over which their expected costs are accumulated. Throughout, Greek letters denote measures on the concrete state space, and Latin letters denote their densities where these exist. Let us then consider the hyper-rectangular state space
\begin{equation}
	\mathcal{X}:=\prod_{i=1}^{n}[a^{(i)},b^{(i)}]\subset\mathbb{R}^{n},
	\label{eq:hyperrectangle}
\end{equation}
with $-\infty<a^{(i)}<b^{(i)}<\infty$ for every dimension $i\in\{1,\dots,n\}$.
We write $\operatorname{diag}(\mathcal{X}):=(\sum_{i=1}^{n}(b^{(i)}-a^{(i)})^2)^{1/2}$ for the Euclidean diagonal length of this hyper-rectangle. Let us further define the concrete continuous-state MC, the subject of the methods developed below, as follows.

\begin{defn}[Continuous-state Markov chain]\label{def:continuous_state_markov_chain}
	A continuous-state MC is a tuple
	\begin{equation*}
		\mathcal{M}_{\mathcal{X}}:=\left(\mathcal{X},\pi_0,c,\tau\right),
	\end{equation*}
	where $\mathcal{X}$ is the state space, $\pi_0\in\mathcal{P}(\mathcal{X})$ is the initial-state measure, so that $\pi_0(A)=\mathbb{P}(x_0\in A)$, $\tau$ is a transition kernel satisfying $\tau(\,\cdot\,|x)\in\mathcal{P}(\mathcal{X})$ for every $x\in\mathcal{X}$, and $c:\mathcal{X}\to\mathbb{R}$ is a bounded measurable stage-cost function. The state process $(x_k)_{k\ge0}$ evolves according to
	\begin{equation*}
		\mathbb{P}(x_{k+1}\in A|x_k=x)=\tau(A|x),
	\end{equation*}
	for every measurable $A\subseteq\mathcal{X}$ and $x\in\mathcal{X}$. We let
	\begin{equation*}
		J(\mathcal{M}_{\mathcal{X}}):=\mathbb{E}_{\mathcal{M}_{\mathcal{X}}}\left[\sum_{k=0}^{H-1}c(x_k)\right],
	\end{equation*}
    denote the expected cumulative cost associated with stage cost $c(x_k)$ w.r.t.\ the distribution of the stochastic state sequence $(x_k)_{k=0}^{H-1}$ induced by $x_0\sim \pi_0$, and $x_{k+1}\sim \tau(\,\cdot\,|x_k)$ for $k\in\{0,\dots,H-1\}$.
\end{defn}

\begin{defn}[Time-varying Markov chain]\label{def:time_varying_markov_chain}
	A time-varying Markov chain (MC) is a tuple
	\begin{equation*}
		\mathcal{M}:=(S,\alpha,c,P),
	\end{equation*}
	where $S$ is its finite state space, $\alpha\in\mathcal{P}(S)$ is the initial distribution, $c:\{0,1,\ldots,H-1\}\times S\to\mathbb{R}$ is the stage-cost function and $P=(P_0,P_1,\dots,P_{H-1})$, where $P_k(s,s'\,)=\mathbb{P}({s_{k+1}=s'}| {s_k=s})$ is the probability at time $k$ of transitioning from state $s$ to state $s'$, and $H\in\mathbb{N}$ is the time horizon. We further let
	\begin{equation*}
		J(\mathcal{M}):=\mathbb{E}_{\mathcal{M}}\left[\sum_{k=0}^{H-1}c_k(s_k)\right],
	\end{equation*}
	denote its expected cumulative cost induced by $s_0\sim\alpha$, and $s_{k+1}\sim P_k(s_k,\,\cdot\,)$ for $k\in\{0,\dots,H-1\}$.
\end{defn}

An IMC specifies closed intervals within which the transition and cost entries of such time-varying MCs are allowed to vary, essentially defining a family of infinitely many time-varying MCs (see e.g.~\cite{Givan2000} for an introduction).

\begin{defn}[Interval Markov chain]\label{def:interval_markov_chain}
	An interval Markov chain (IMC) is a tuple
	\begin{equation*}
		\mathcal{I}:=\left(S,\alpha,\underline{c},\overline{c},\underline{P},\overline{P}\right),
	\end{equation*}
	where $S$ and $\alpha$ are as in Definition~\ref{def:time_varying_markov_chain}, while $\underline{c},\overline{c}:\{0,1,\ldots,H-1\}\times S\to\mathbb{R}$ are such that $\underline{c}_k(s)\leq\overline{c}_k(s)$, for every $s\in S$, and every $k\in\{0,\dots,H-1\}$. Then $[\underline{c}_k(s),\overline{c}_k(s)]$ is the admissible stage-cost and $[\underline{P}_k(s,s'),\overline{P}_k(s,s')]$ is the admissible transition-probability interval at $k\in\{0,\dots,H-1\}$ at $s\in S$. The transition intervals at state $s$, time $k$, define the nonempty set
	\begin{equation*}
		\mathsf{P}_{k}(s):=\left\{p\in\mathcal{P}(S)\mathrel{\big|}\underline{P}_k(s,s')\le p(s')\le\overline{P}_k(s,s'),\ s'\in S\right\}.
	\end{equation*}
\end{defn}

We interpret an IMC $\mathcal{I}$ as representing the collection of all time-varying MCs $\mathcal{M}$ with initial distribution $\alpha$, with $c_k(s)\in[\underline{c}_k(s),\overline{c}_k(s)]$ for every $k\in\{0,\dots,H-1\}$, and $P_k(s,\,\cdot\,)\in\mathsf{P}_k(s)$ for every $k\in\{0,\dots,H-1\}$. With slight abuse of notation, we write $\mathcal{M}\in\mathcal{I}$ for membership in this collection. The values of the minimum and maximum expected cumulative costs that can be found amongst the MCs $\mathcal{M}\in\mathcal{I}$ can be obtained by robust dynamic programming~\cite{Givan2000}.

\begin{defn}[Robust expected-cost bounds]\label{def:robust_expected_cost_bounds}
	Let $\underline{V}_H(s)=\overline{V}_H(s)=0$. For $k\in\{H-1,\dots,0\}$, define
	\begin{equation}
	\begin{split}
		\underline{V}_{k}(s)&:=\underline{c}_{k}(s)+\min_{p\in\mathsf{P}_k(s)}\sum_{s'\in S}p(s')\underline{V}_{k+1}(s'),\\
		\overline{V}_{k}(s)&:=\overline{c}_{k}(s)+\max_{p\in\mathsf{P}_k(s)}\sum_{s'\in S}p(s')\overline{V}_{k+1}(s').
	\end{split}
		\label{eq:robust_bellman_recursion}
	\end{equation}
	The \emph{robust expected-cost bounds} associated with the IMC $\mathcal{I}$ are then
	\begin{equation}
		\underline{J}(\mathcal{I}):=\sum_{s\in S}\alpha(s)\underline{V}_0(s), \quad \overline{J}(\mathcal{I}):=\sum_{s\in S}\alpha(s)\overline{V}_0(s).
		\label{eq:robust_performance_bounds}
	\end{equation}
\end{defn}

\begin{prop}[Robust expected-cost extrema]
For any IMC $\mathcal{I}$, the robust bounds in~\eqref{eq:robust_performance_bounds} satisfy
\begin{equation}
	\underline{J}(\mathcal{I})=\min_{\mathcal{M}\in\mathcal{I}}J(\mathcal{M}), \quad \overline{J}(\mathcal{I})=\max_{\mathcal{M}\in\mathcal{I}}J(\mathcal{M}).
	\label{eq:imc_member_cost_extrema}
\end{equation}
\end{prop}
This proposition uses the standard interval dynamic-programming characterization of extrema over bounded-parameter Markov models; see, e.g., \cite{Givan2000}.

\subsection{The abstraction problem}\label{sec:abstractionproblem}

Given a continuous-state Markov chain $\mathcal{M}_{\mathcal{X}}$ as in Definition~\ref{def:continuous_state_markov_chain}, the \textit{abstraction problem} is to construct a finite-state IMC abstraction $\mathcal{I}$ whose robust expected-cost bounds contain the expected cumulative cost of the concrete system. We call this property \emph{expected-cost soundness}. Among sound abstractions, the robust expected-cost bounds should satisfy
\begin{equation*}
	\underline{J}(\mathcal{I})\leq J(\mathcal{M}_{\mathcal{X}})\leq\overline{J}(\mathcal{I}).
\end{equation*}
Among sound abstractions, tighter bounds are preferable. We measure the tightness of the resulting expected-cost interval by its width $\overline{J}(\mathcal{I})-\underline{J}(\mathcal{I})$. In Section~\ref{sec:memory_dependent_imcs}, we construct sound IMC abstractions whose abstract states remember the regions of the state space visited in the recent past. Section~\ref{sec:refinement_efficiency} then analyzes how memory lengthening and partition refinement affect the resulting bounds.

\section{Memory-dependent and no-memory IMC abstractions}\label{sec:memory_dependent_imcs}

This section constructs memory-dependent and no-memory IMC abstractions of $\mathcal{M}_{\mathcal{X}}$ and proves that they are sound. We then conclude by showing that adding memory to an existing IMC abstraction can only tighten the resulting cost and transition intervals. The proofs of the results are given in Appendix~\ref{sec:proofs}.

Assume without loss of generality that $\mathcal{X}$ carries a uniform hyper-rectangular partition: for a \emph{base-resolution parameter} $M\in\mathbb{N}_{\ge1}$, each coordinate interval $[a^{(i)},b^{(i)}]$ in~\eqref{eq:hyperrectangle} is split uniformly into $M$ subintervals, their Cartesian product forming the \emph{base partition} $\mathcal{Y}:=\{y_1,\ldots,y_{M^n}\}$. For a \emph{refinement factor} $N\ge1$, splitting each cell of $\mathcal{Y}$ uniformly into $N^{n}$ parts yields the \emph{refined partition} $\mathcal{R}:=\{r_1,\ldots,r_{(MN)^n}\}$, so that every refined cell $r\in\mathcal{R}$ lies in a unique parent cell $y\in\mathcal{Y}$.

Let an abstract state retain the current refined cell $r_k$ and the $\ell$ preceding ones, with $\ell\ge0$ referred to as the \emph{memory length}. This yields a \emph{memory state} $m_k:=(r_{k-\ell},\ldots,r_k)\in\mathcal{R}^{\ell+1}$; for $k<\ell$ the history is shorter, and the effective memory is $m_k:=(r_0,\ldots,r_k)$. Hence $N$ controls the partition resolution, and $\ell$ the length of the retained past.

\subsection{Memory-dependent IMC abstractions}\label{sec:memory_dependent_imc_abstractions}

Disregarding cell shape or memory length, let $\mathsf{M}:=\bigcup_{\ell=0}^{\infty}(\mathcal{B}(\mathcal{X}))^{\ell+1}$ describe the set of all possible finite-length (Borel) cell-visitation memories that could exist over time in a concrete state space $\mathcal{X}$. We identify a one-cell memory state $m_k=(r_k)$ with its cell $m_k=r_k$. A pair of memory states $m_k,m_{k+1}\in\mathsf{M}$, both of length $\ell+1$, for $\ell\ge 0$, is called \emph{shift-compatible} when
\begin{equation*}
	\begin{split}
		m_{k}=(r_{k-\ell},&r_{k-\ell+1},r_{k-\ell+2},\ldots,r_k), \\
		m_{k+1}=(&r_{k-\ell+1},r_{k-\ell+2},\ldots,r_k,r_{k+1}).
	\end{split}
\end{equation*}
This property is similar to the so-called \emph{domino game} in \cite{SchmuckRaisch2014}. A shift-compatible successor $m_{k+1}$ can thus be constructed from $m_{k}$ by discarding its oldest cell and appending a new current cell $r_{k+1}$. Let the tail (every remembered cell except the oldest) of such a memory state $m_k$ define the event $E_{m_k}$, as
\begin{equation*}
	E_{m_k}:=\{(x_{k-\ell+1},x_{k-\ell+2},\dots,x_{k})\in r_{k-\ell+1}\times\hspace{-0.5mm}\cdots\hspace{-0.5mm}\times r_{k}\}.
\end{equation*}
For $\ell=0$, this is set to the certain event by convention. 

We proceed to introduce the foundational object of memory-dependent IMC abstractions, namely the probability measure over the current cell $r_k$, as conditioned on the memory tail event $E_{m_k}$ and a concrete starting state $\xi\in r_{k-\ell}$. For a memory state $m_k=(r_{k-\ell},\ldots,r_k)\in\mathsf{M}$, a concrete state $\xi\in r_{k-\ell}$ is called \emph{admissible} if
$\mathbb{P}(E_{m_k}|x_{k-\ell}=\xi)>0$. For every admissible $\xi$, and every $A\in\mathcal{B}(r_k)$, define the memory-conditioned measure over the current cell $\mu_{\xi,m_k}^{(\ell)}\in\mathcal{P}(r_k)$, as
\begin{equation}\label{eq:memory_conditional_measure}
	\mu_{\xi,m_k}^{(\ell)}(A):=\mathbb{P}(x_k\in A|E_{m_k},x_{k-\ell}=\xi)=\frac{\mathbb{P}(x_k\in A,E_{m_k}|x_{k-\ell}=\xi)}{\mathbb{P}(E_{m_k}|x_{k-\ell}=\xi)}.
\end{equation}
For $\ell=0$, this yields $\mu_{\xi,m_k}^{(0)}=\delta_\xi$. 

\begin{prop}[Recursivity]\label{prop:recursivity}
	Fix a memory state $m_k=(r_{k-\ell},\ldots,r_k)$, and an admissible $\xi\in r_{k-\ell}$. For every measurable $A\subseteq\mathcal{X}$, initialize with $\beta_{k-\ell}^{\xi,m_k}(A):=\delta_\xi(A)$, and, for $i=k-\ell,\ldots,k-1$, recursively define
	\begin{equation*}
		\beta_{i+1}^{\xi,m_k}(A):=\int_{r_i}\tau(A\cap r_{i+1}|x)\,\beta_i^{\xi,m_k}(\mathrm{d}x).
	\end{equation*}
	Then
	\begin{equation}
	\mu_{\xi,m_k}^{(\ell)}(A)=\frac{\beta_k^{\xi,m_k}(A)}{\beta_k^{\xi,m_k}(r_k)}.
	\label{eq:normalization_at_the_end}
	\end{equation}
\end{prop}

Let, for every finite memory state $m_k=(r_{k-\ell},\ldots,r_k)\in\mathsf{M}$ for which at least one admissible $\xi\in r_{k-\ell}$ exists, the memory-conditioned stage-cost interval functions be defined as
\begin{equation}
\begin{split}
	\underline{c}(m_k)&:=\hspace{-2mm}\inf_{\substack{\xi\in r_{k-\ell\hspace{2mm}} \\
	\mathbb{P}(E_{m_k}|x_{k-\ell}=\xi)>0}}\hspace{0mm}\int_{r_k}\hspace{-2mm}c(x)\mu_{\xi,m_k}^{(\ell)}(\mathrm{d}x),\\
	\overline{c}(m_k)&:=\hspace{-2mm}\sup_{\substack{\xi\in r_{k-\ell\hspace{2mm}} \\
	\mathbb{P}(E_{m_k}|x_{k-\ell}=\xi)>0}}\hspace{0mm}\int_{r_k}\hspace{-2mm}c(x)\mu_{\xi,m_k}^{(\ell)}(\mathrm{d}x).
\end{split}
	\label{eq:memory_cost_bounds}
\end{equation}
For every shift-compatible pair $(m_k,m_{k+1})\in\mathsf{M}\times\mathsf{M}$, define the memory-conditioned transition-probability interval functions by
\begin{equation}
\begin{split}
	&\underline{P}(m_k,m_{k+1}):=\hspace{-2mm}\inf_{\substack{\xi\in r_{k-\ell} \\
	\mathbb{P}(E_{m_k}|x_{k-\ell}=\xi)>0}}\hspace{-1mm}\int_{r_k}\tau(r_{k+1}|x)\mu_{\xi,m_k}^{(\ell)}(\mathrm{d}x),\\
	&\overline{P}(m_k,m_{k+1}):=\hspace{-2mm}\sup_{\substack{\xi\in r_{k-\ell} \\
	\mathbb{P}(E_{m_k}|x_{k-\ell}=\xi)>0}}\hspace{-1mm}\int_{r_k}\tau(r_{k+1}|x)\mu_{\xi,m_k}^{(\ell)}(\mathrm{d}x).
\end{split}
	\label{eq:memory_transition_intervals}
\end{equation}
If no admissible $\xi\in r_{k-\ell}$ exists for a finite memory state $m_k$, we set the intervals in~\eqref{eq:memory_cost_bounds} to the corresponding one-cell intervals, $\underline{c}(m_k):=\underline{c}(r_k)$, $\overline{c}(m_k):=\overline{c}(r_k)$, and for every shift-compatible successor $m_{k+1}$, we likewise set $\underline{P}(m_k,m_{k+1}):=\underline{P}(r_k,r_{k+1})$, $\overline{P}(m_k,m_{k+1}):=\overline{P}(r_k,r_{k+1})$. Note that such $m_k$ are unreachable from every concrete state in its oldest cell and will thus never be reached, nonetheless this interval selection ensures soundness even when this is disregarded. For every pair $(m_k,m_{k+1})$ that is not shift-compatible, we further have
\begin{equation*}
\underline P(m_k,m_{k+1})=\overline P(m_k,m_{k+1})=0.
\end{equation*}

\begin{figure}[tb]
\centering
\import{vis/}{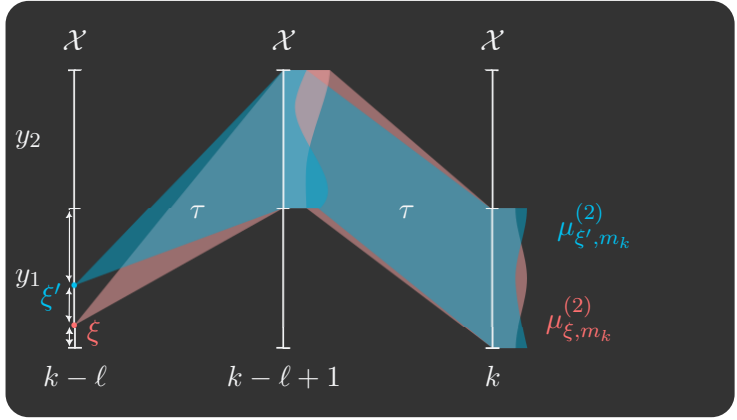}
\caption{A visualisation of how the choice of $\xi$ at time $k-\ell$ affects the measure $\mu_{\xi,m_k}^{(2)}$ at time $k$, subject to $m_{k}=(y_1,y_2,y_1)$.}
\label{fig:memory_method}
\end{figure}

Figure~\ref{fig:memory_method} illustrates how the selection of $\xi$ affects $\mu_{\xi,m_k}^{(\ell)}$; the fundamental mechanism behind optimizations~\eqref{eq:memory_cost_bounds} and~\eqref{eq:memory_transition_intervals}.

\def\svgwidth{0.57\columnwidth} % this defines the scaling of the image

\begin{defn}[Memory-dependent IMC abstraction]\label{def:memory_dependent_imc_abstraction}
	The memory-dependent IMC abstraction of $\mathcal{M}_{\mathcal{X}}$ associated with the refined partition $\mathcal{R}$ and memory length $\ell$ is the tuple
	\begin{equation*}
		\mathcal{I}^{(\ell)}:=\left(S^{(\ell)},\alpha^{(\ell)},\underline{c}^{(\ell)},\overline{c}^{(\ell)},\underline{P}^{(\ell)},\overline{P}^{(\ell)}\right),
	\end{equation*}
	where $S^{(\ell)}:=\mathcal{R}^{\ell+1}$ is the finite-memory state space, and $\alpha^{(\ell)}\in\mathcal{P}(S^{(\ell)})$ is the initial distribution with $\alpha^{(\ell)}(m_0):=\frac{1}{|\mathcal{R}|^\ell}\pi_0(r_0)$, for all $m_0\in S^{(\ell)}$. For $k\in\{0,1,\dots,H-1\}$ where $k<\ell$ and $m_k=(r_{k-\ell},\ldots,r_k)$, let the available suffix of $m_k$ be $\sigma_k(m_k):=(r_{0},\ldots,r_{k})$. For every finite memory state $m_k\in S^{(\ell)}$, define the time-$k$ cost intervals as
	\begin{equation*}
		\begin{split}
			\underline{c}_k^{(\ell)}(m_k):=\begin{cases}
				\underline{c}(\sigma_k(m_k)),&\text{if }k<\ell,\\
				\underline{c}(m_k),&\text{if }k\ge\ell,
			\end{cases} \\
			\overline{c}_k^{(\ell)}(m_k):=\begin{cases}
				\overline{c}(\sigma_k(m_k)),&\text{if }k<\ell,\\
				\overline{c}(m_k),&\text{if }k\ge\ell,
			\end{cases}
		\end{split}
	\end{equation*}
	and for every $m_{k+1}\in S^{(\ell)}$, the transition-probability intervals through
	\begin{equation*}
		\begin{split}
			\underline{P}_k^{(\ell)}(m_k,m_{k+1}):=\begin{cases}
				\frac{\underline{P}(\sigma_k(m_k),\sigma_{k}(m_{k+1}))}{|\mathcal{R}|^{\ell-k-1}},&\text{if }k<\ell,\\
				\underline{P}(m_k,m_{k+1}),&\text{if }k\ge\ell,
			\end{cases} \\
			\overline{P}_k^{(\ell)}(m_k,m_{k+1}):=\begin{cases}
				\frac{\overline{P}(\sigma_k(m_k),\sigma_{k}(m_{k+1}))}{|\mathcal{R}|^{\ell-k-1}},&\text{if }k<\ell,\\
				\overline{P}(m_k,m_{k+1}),&\text{if }k\ge\ell.
			\end{cases}
		\end{split}
	\end{equation*}
	The correction factor $1/|\mathcal{R}|^{\ell-k-1}$ distributes the transition probability associated with a successor $\sigma_{k+1}(m_{k+1})$, when $k+1\le \ell$, over all equivalent full-memory states $\{m'_{k+1}\in S^{(\ell)} : \sigma_{k+1}(m'_{k+1}) = \sigma_{k+1}(m_{k+1})\}$. 
\end{defn}

\begin{rem}[Classic IMC as a special case] \label{rem:specialcaseclassic}
For $\ell=0$, the Definition~\ref{def:memory_dependent_imc_abstraction} construction reduces to the classic IMC construction induced by $\mathcal{R}$ known from the literature~\cite{Givan2000}.
\end{rem}

\begin{figure}[tb]
	\centering
\import{vis/}{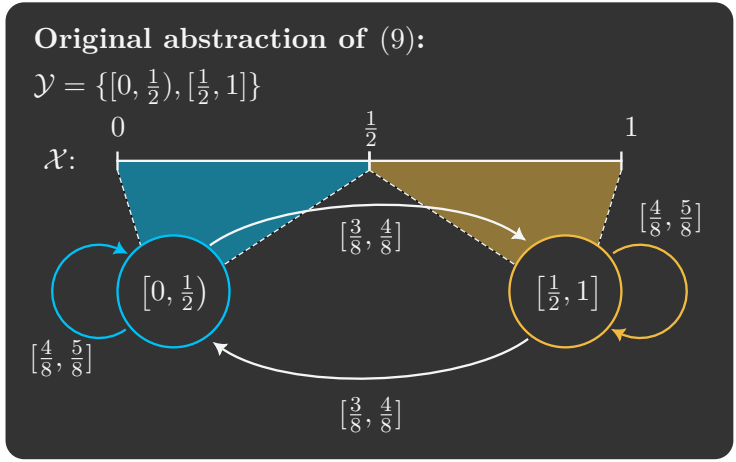}
	\caption{The classic IMC abstraction of~\eqref{eq:example_system}, with $\ell=0$, $N=1$, and $M=2$.}
	\label{fig:basis_system}
\end{figure}

\begin{figure}[tb]
	\centering
\import{vis/}{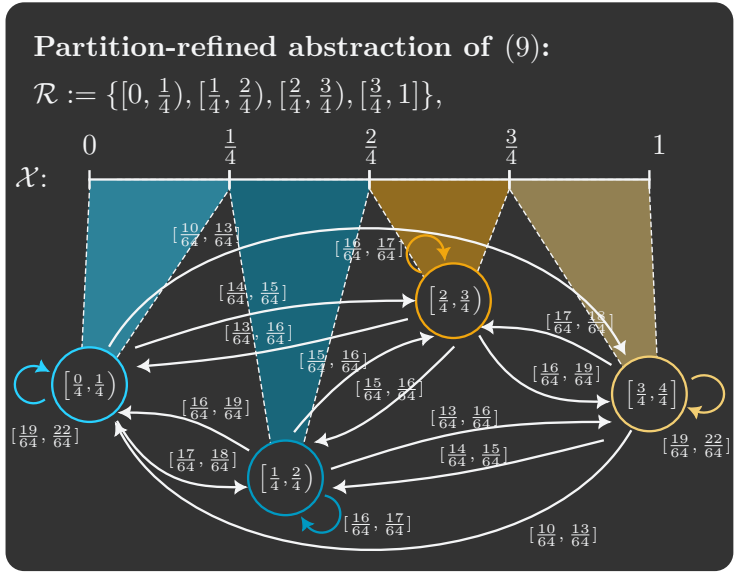}
	\caption{The partition-refined abstraction of~\eqref{eq:example_system}, with $\ell=0$, $N=2$, and $M=2$.}
	\label{fig:partition_refinement}
\end{figure}

\begin{figure}[tb]
	\centering
\import{vis/}{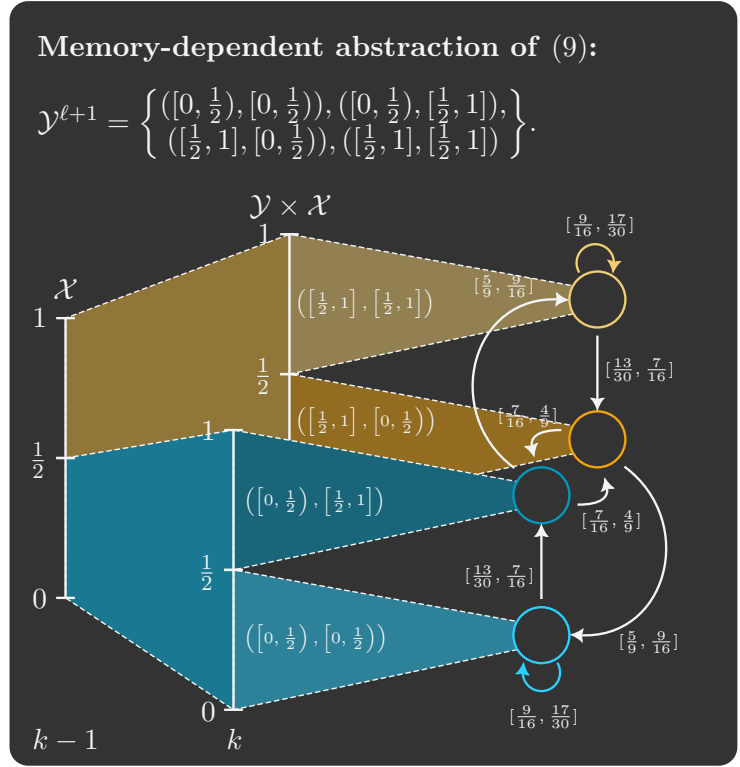}
	\caption{The memory-dependent IMC abstraction of~\eqref{eq:example_system}, with memory length $\ell=1$, $N=1$, and $M=2$.}
	\label{fig:memory_dependent_abstraction}
\end{figure}

\begin{exmp}[A one-dimensional example system]\label{ex:running-example}
Let us introduce a simple example system to illustrate some of the relevant concepts visually. Consider a one-dimensional Markov chain $\mathcal{M}_{\mathcal{X}}$ with state space $\mathcal{X}:=[0,1]$, stage cost $c(x_k):=x_k$, and for any $A\in\mathcal{B}(\mathcal{X})$, initial distribution $\pi_0(A):=\lambda(A)$, with $\lambda$ being the Lebesgue measure, and transition kernel
\begin{equation}
	\tau( A|x_{k}) := \int_{A}(2x_{k}-1)x_{k+1}+(1.5-x_k)~\mathrm{d}x_{k+1}.
	\label{eq:example_system}
\end{equation}
Abstracting this system uniformly with $M=2$ gives the base partition $\mathcal{Y}=\{[0,0.5),[0.5,1]\}$. The associated $N=1$, $\ell=0$, classic IMC abstraction is visualized in Figure~\ref{fig:basis_system}. Following Definition~\ref{def:memory_dependent_imc_abstraction}, Figure~\ref{fig:partition_refinement} visualizes the classic IMC abstraction with $\ell=0$ and partition refinement factor $N=2$. Figure~\ref{fig:memory_dependent_abstraction} shows its memory-dependent IMC with memory length one and partition refinement factor $N=1$.
\end{exmp}

\def\svgwidth{0.57\columnwidth} % this defines the scaling of the image

\def\svgwidth{0.57\columnwidth} % this defines the scaling of the image

\def\svgwidth{0.57\columnwidth} % this defines the scaling of the image

The following is the first main result of the present paper, showing that memory-dependent abstractions are sound.

\begin{thm}[Expected-cost soundness of memory-dependent abstractions]\label{thm:memory_dependent_soundness}
	For every memory length $\ell\in\mathbb{N}_{\ge0}$, the memory-dependent IMC abstraction $\mathcal{I}^{(\ell)}$ in Definition~\ref{def:memory_dependent_imc_abstraction} is guaranteed to satisfy
	\begin{equation}
		\underline{J}(\mathcal{I}^{(\ell)})\le J(\mathcal{M}_{\mathcal{X}})\le\overline{J}(\mathcal{I}^{(\ell)}).
		\label{eq:memory_dependent_soundness}
	\end{equation}
\end{thm}

We further present a local interval-inclusion result, concluding that a memory-conditioned interval is always contained within the corresponding classic interval of the current cell.

\begin{lem}[Memory conditioning cannot widen local intervals]\label{lem:memory_cellwise_interval_inclusion}
	For every $\ell\in\mathbb{N}_{\ge0}$, time $k\ge\ell$, memory state $m_k=(r_{k-\ell},\ldots,r_k)\in\mathcal{R}^{\ell+1}$, and shift-compatible successor $m_{k+1}=(r_{k-\ell+1},\ldots,r_k,r_{k+1})\in\mathcal{R}^{\ell+1}$,
	\begin{equation}
		\underline{c}(r_k)\le\underline{c}(m_k)\le\overline{c}(m_k)\le\overline{c}(r_k),
		\label{eq:memory_cost_inclusion}
	\end{equation}
	and
	\begin{equation}
		\underline{P}(r_k,r_{k+1})\le\underline{P}(m_k,m_{k+1})\le\overline{P}(m_k,m_{k+1})\le\overline{P}(r_k,r_{k+1}).
		\label{eq:memory_transition_inclusion}
	\end{equation}
\end{lem}

\subsection{No-memory abstractions}\label{sec:no_memory_imc}

For the sake of simplicity, consider the memory-dependent construction at $\ell=1$. The cost and transition intervals at $m_k$ are then obtained by conditioning on, and optimizing over $x_{k-1}=\xi\in r_{k-1}$. Let us now forget the previous abstract cell $r_{k-1}$. In the memory-dependent construction, this yields an $\ell=0$ system, defaulting to the classic IMC abstraction. In this section we consider what happens if we alternatively follow the $\ell=1$ construction with the $\xi\in r_{k-1}$ constraint relaxed. More generally, what happens if for any $\ell\in\mathbb{N}$, we no longer condition on $E_m$, and we relax the constraint $\xi\in r_{k-\ell}$, no longer requiring memorization of $r_{k-\ell},r_{k-\ell+1},\dots,r_{k-1}$ anymore? This yields a no-memory abstraction, an abstraction whose state space remains $\mathcal{R}$ while still able to exploit more transition kernel information than a classic IMC abstraction. As, in contrast to classic IMC abstractions, the no-memory abstraction is not a special case of the memory-dependent abstraction, we describe it separately here.

For $\ell\in\mathbb{N}$, $\xi\in\mathcal{X}$, and $r_k\in\mathcal{R}$ satisfying
$\mathbb{P}(x_k\in r_k|x_{k-\ell}=\xi)>0$, define the no-memory conditional measure
$\mu_{\xi,r_k}^{(\ell),\mathrm{n}}\in\mathcal{P}(r_k)$ as the no-memory equivalent of~\eqref{eq:memory_conditional_measure}, as $\mu_{\xi,r_k}^{(\ell),\mathrm{n}}(A):=\mathbb{P}(x_k\in A|x_{k-\ell}=\xi,\ x_k\in r_k)$. For every $\ell\in\mathbb{N}$ and current refined cell $r_k\in\mathcal{R}$ for which at least one $\xi\in\mathcal{X}$ satisfies $\mathbb{P}(x_k\in r_k|x_{k-\ell}=\xi)>0$, define the no-memory stage-cost interval functions as
\begin{equation*}
\begin{split}
	\underline{c}^{(\ell),\mathrm{n}}(r_k)&:=\hspace{-3mm}\inf_{\substack{\xi\in\mathcal{X} \\
	\mathbb{P}(x_k\in r_k|x_{k-\ell}=\xi)>0}}\hspace{-3mm}\int_{r_k}c(x)\,\mu_{\xi,r_k}^{(\ell),\mathrm{n}}(\mathrm{d}x),\\
	\overline{c}^{(\ell),\mathrm{n}}(r_k)&:=\hspace{-3mm}\sup_{\substack{\xi\in\mathcal{X} \\
	\mathbb{P}(x_k\in r_k|x_{k-\ell}=\xi)>0}}\hspace{-3mm}\int_{r_k}c(x)\,\mu_{\xi,r_k}^{(\ell),\mathrm{n}}(\mathrm{d}x).
\end{split}
\end{equation*}

For every $r_k,r_{k+1}\in\mathcal{R}$, define the no-memory transition-probability interval functions as
\begin{equation*}
\begin{split}
	\underline{P}^{(\ell),\mathrm{n}}(r_k,r_{k+1})&:=\hspace{-7mm}\inf_{\substack{\xi\in\mathcal{X} \\
	\mathbb{P}(x_k\in r_k|x_{k-\ell}=\xi)>0}}\hspace{-3mm}\int_{r_k}\tau(r_{k+1}|x)\,\mu_{\xi,r_k}^{(\ell),\mathrm{n}}(\mathrm{d}x),\\
	\overline{P}^{(\ell),\mathrm{n}}(r_k,r_{k+1})&:=\hspace{-7mm}\sup_{\substack{\xi\in\mathcal{X} \\
	\mathbb{P}(x_k\in r_k|x_{k-\ell}=\xi)>0}}\hspace{-3mm}\int_{r_k}\tau(r_{k+1}|x)\,\mu_{\xi,r_k}^{(\ell),\mathrm{n}}(\mathrm{d}x).
\end{split}
\end{equation*}

\begin{defn}[No-memory IMC abstraction]\label{def:no_memory_imc_abstraction}
	The no-memory interval Markov chain (IMC) abstraction of $\mathcal{M}_{\mathcal{X}}$ associated with the refined partition $\mathcal{R}$ and memory length $\ell$ is the time-varying tuple
	\begin{equation*}
		\mathcal{I}^{(\ell),\mathrm{n}}:=\left(S^{\mathrm{n}},\alpha^{\mathrm{n}},\underline{c}^{(\ell),\mathrm{n}},\overline{c}^{(\ell),\mathrm{n}},\underline{P}^{(\ell),\mathrm{n}},\overline{P}^{(\ell),\mathrm{n}}\right),
	\end{equation*}
	where $S^{\mathrm{n}}:=\mathcal{R}$ is the finite state space, and $\alpha^{\mathrm{n}}\in\mathcal{P}(\mathcal{R})$ is the initial distribution defined by $\alpha^{\mathrm{n}}(r_0):=\pi_0(r_0)$. For every $k\in\{0,1,\dots,H-1\}$, and $r_k,r_{k+1}\in\mathcal{R}$, define the time-$k$ interval functions as
	\begin{equation*}
	\begin{split}
		\underline{c}^{(\ell),\mathrm{n}}_k(r_k)&:=\begin{cases}
			\underline{c}^{(k),\mathrm{n}}(r_k),&\text{if }0\le k<\ell,\\
			\underline{c}^{(\ell),\mathrm{n}}(r_k),&\text{if }k\ge\ell,
		\end{cases}\\
		\overline{c}^{(\ell),\mathrm{n}}_k(r_k)&:=\begin{cases}
			\overline{c}^{(k),\mathrm{n}}(r_k),&\text{if }0\le k<\ell,\\
			\overline{c}^{(\ell),\mathrm{n}}(r_k),&\text{if }k\ge\ell,
		\end{cases}
	\end{split}
	\end{equation*}
	and
	\begin{equation*}
	\begin{split}
		\underline{P}^{(\ell),\mathrm{n}}_k(r_k,r_{k+1})&:=\begin{cases}
			\underline{P}^{(k),\mathrm{n}}(r_k,r_{k+1}),&\text{if }0\le k<\ell,\\
			\underline{P}^{(\ell),\mathrm{n}}(r_k,r_{k+1}),&\text{if }k\ge\ell,
		\end{cases}\\
		\overline{P}^{(\ell),\mathrm{n}}_k(r_k,r_{k+1})&:=\begin{cases}
			\overline{P}^{(k),\mathrm{n}}(r_k,r_{k+1}),&\text{if }0\le k<\ell,\\
			\overline{P}^{(\ell),\mathrm{n}}(r_k,r_{k+1}),&\text{if }k\ge\ell.
		\end{cases}
	\end{split}
	\end{equation*}
	For all $r_k\in \mathcal{R}$ and $\ell\ge 1$ such that $\mathbb{P}(x_k\in r_k|x_{k-\ell}=\xi)=0$ for all $\xi\in\mathcal{X}$, use $\underline{c}^{(0),\mathrm{n}}(r_k)$, $\overline{c}^{(0),\mathrm{n}}(r_k)$, $\underline{P}^{(0),\mathrm{n}}(r_k,r_{k+1})$, and $\overline{P}^{(0),\mathrm{n}}(r_k,r_{k+1})$.
\end{defn}

For $\ell=1$, this recovers the no-memory construction described at the beginning of this subsection. The mixture argument in Theorem~\ref{thm:memory_dependent_soundness} applies equivalently to this new construction, giving
\begin{equation*}
	\underline{J}(\mathcal{I}^{(\ell),\mathrm{n}})\le J(\mathcal{M}_{\mathcal{X}})\le\overline{J}(\mathcal{I}^{(\ell),\mathrm{n}}).
\end{equation*}

\section{Bounding the robust value-function widths}\label{sec:refinement_efficiency}

From the preceding section, we know that memory conditioning yields nested local intervals. This section takes this one step further by deriving analytical upper bounds on the robust value-function width as functions of the memory length $\ell$ and partition refinement factor $N$. The analytical nature of these bounds enables identification of properties of the concrete continuous MC that affect the rate of tightening. The number of effective state transitions in an abstraction model (which governs the robust dynamic-programming complexity) is employed to quantify model complexity below.

\subsection{Bounding the general robust value-function width}\label{sec:filter_forgetting_value_function_widths}

Let $\mathcal{Y}$ be the uniform $M$-grid of~\eqref{eq:hyperrectangle}, and write $d_M:=\operatorname{diag}(\mathcal{X})/M$ for its cell diameter. The memory-dependent bound considers that base grid, i.e., $N=1$, identifying $\mathcal{R}=\mathcal{Y}$, while the partition-refinement bound treats $N$ separately. Let $c_{\min}\le c(x)\le c_{\max}$ for all $x\in\mathcal{X}$, and write $C_c:=c_{\max}-c_{\min}$.

For probability measures $\mu,\nu$ on $\mathcal{X}$, their 1-Wasserstein distance~\cite{Villani2009OptimalTransport} is
\begin{equation}
	W_1(\mu,\nu):=\inf_{\gamma\in\Gamma(\mu,\nu)}\int\|x-\tilde{x}\|_2\,\gamma(\mathrm{d}x,\mathrm{d}\tilde{x}),
	\label{eq:wasserstein_distance}
\end{equation}
where $\Gamma(\mu,\nu)$ is the set of their couplings.

\begin{defn}[Remaining distributional ambiguity]\label{def:memory_conditioned_dispersion}
	For $\ell\in\mathbb{N}_{\ge1}$ and $m\in\mathcal{Y}^{\ell+1}$ for which at least one admissible $\xi\in y_{-\ell}$ exists, define the remaining distributional ambiguity of $m$ as
	\begin{equation*}
		\Delta_\ell(m):=\sup_{\xi,\zeta}W_1\bigl(\mu_{\xi,m}^{(\ell)},\mu_{\zeta,m}^{(\ell)}\bigr),
	\end{equation*}
	where the supremum ranges over $\xi,\zeta\in y_{-\ell}$ satisfying $\mathbb{P}(E_m|x_{-\ell}=\xi)>0$ and $\mathbb{P}(E_m|x_{-\ell}=\zeta)>0$. For a memory state $m\in\mathcal{Y}^{\ell+1}$ with no admissible $\xi\in y_{-\ell}$, set $\Delta_\ell(m):=0$. The worst-case remaining distributional ambiguity at memory length $\ell$ is $\Delta_\ell:=\max_{m\in\mathcal{Y}^{\ell+1}}\Delta_\ell(m)$.
\end{defn}

Thus, $\Delta_\ell$ is the worst-case 1-Wasserstein distance between two memory-conditioned distributions $\mu_{\xi,m}^{(\ell)}$ and $\mu_{\zeta,m}^{(\ell)}$, taken over all $m\in\mathcal{Y}^{\ell+1}$ and admissible $\xi,\zeta\in y_{-\ell}$. To control transition-probability interval widths, we impose the following cellwise sensitivity assumption on the transition density.

\begin{assum}[Cellwise transition-density sensitivity]\label{ass:cellwise_transition_density_sensitivity}
    The transition kernel $\tau$ admits a strictly positive, continuous density $t$ on $\mathcal{X}\times\mathcal{X}$, so that $\tau(A|x)=\int_A t(x'|x)\,\mathrm{d}x'$, for every measurable $A\subseteq\mathcal{X}$ and $x\in\mathcal{X}$. For every $y\in\mathcal{Y}$, the dependence of $t(x'|x)$ on the departure state $x\in y$ is pointwise Lipschitz with an integrable bound $g_y:\mathcal{X}\to\mathbb{R}_{\ge0}$, as
    \begin{equation}
        |t(x'|x)-t(x'|\tilde{x})|\le g_y(x')\|x-\tilde{x}\|_2,
        \label{eq:cellwise_transition_density_sensitivity}
    \end{equation}
    for all $x,\tilde{x}\in y$ and almost every $x'\in\mathcal{X}$.
\end{assum}

Assumption~\ref{ass:cellwise_transition_density_sensitivity} is an integrable, pointwise-in-\(x'\) refinement of the cellwise Lipschitz conditions used in other abstraction literature~\cite{Soudjani2013Adaptive}, with $G_y$ below aggregating these local constants.

Let	$G_y:=\int_{\mathcal{X}}g_y(x')\,\mathrm{d}x'$, and $G:=\max_{y\in\mathcal{Y}}G_y$. The following is the second main result of the paper, providing --- under mild assumptions --- bounds on robust value-function width as a function of either (i) the memory length $\ell$ or (ii) the partition refinement factor $N$.

\begin{thm}[Robust value-function width upper bounds]\label{thm:absorbing_value_function_width_convergence}
	Fix $\ell\in\mathbb{N}_{\ge1}$. Suppose that Assumption~\ref{ass:cellwise_transition_density_sensitivity} holds and that $c$ is $L_c$-Lipschitz on $\mathcal{X}$. Let $\underline{V}_k^{(\ell)}(m)$ and $\overline{V}_k^{(\ell)}(m)$ be the robust cost-to-go bounds of the memory-dependent abstraction on $m\in\mathcal{Y}^{\ell+1}$, and let $\underline{V}_k^{\mathrm{pr}}(r)$ and $\overline{V}_k^{\mathrm{pr}}(r)$ be the classic IMC abstraction bounds under a uniform partition refinement factor $N$. Define
	\begin{equation*}
	\begin{split}
		D_k^{(\ell)}&:=\max_{m\in\mathcal{Y}^{\ell+1}}\left(\overline{V}_k^{(\ell)}(m)-\underline{V}_k^{(\ell)}(m)\right),\\
		D_k^{\mathrm{pr}}&:=\max_{r\in\mathcal{R}}\left(\overline{V}_k^{\mathrm{pr}}(r)-\underline{V}_k^{\mathrm{pr}}(r)\right).
	\end{split}
	\end{equation*}
	For an $H$-stage recursion evaluated at a time $k\in\{\ell,\ldots,H\}$,
	\begin{equation}
		D_k^{(\ell)}\le(H-k)(L_c+G(H-k-1)C_c/4)\Delta_\ell,
		\label{eq:memory_global_value_function_width}
	\end{equation}
	while
	\begin{equation}
		D_k^{\mathrm{pr}}\le(H-k)(L_c+G(H-k-1)C_c/4)\frac{d_M}{N}.
		\label{eq:partition_global_value_function_width}
	\end{equation}
\end{thm}

The proof of Theorem~\ref{thm:absorbing_value_function_width_convergence} uses two width upper-bounds, one on the robust stage costs and one on the robust transition probability intervals. In both cases, the bounded quantity is the comparison of a corresponding functional under different memory-conditioned measures $\mu_{\xi,m}^{(\ell)}$ and $\mu_{\zeta,m}^{(\ell)}$. As the cost functional is $L_c$-Lipschitz, while the transition density is Lipschitz by Assumption~\ref{ass:cellwise_transition_density_sensitivity}, by Kantorovich--Rubinstein duality, both differences are exactly controlled by the Wasserstein distance between the two measures.

These separate upper bounds independently establish the tightening potential of memory lengthening and partition refinement after the full memory length is available, but neither enable a general ordering of the associated widths.

\begin{prop}[Monotonicity of the remaining distributional ambiguity]\label{prop:dispersion_monotone}
	Memory lengthening is guaranteed to effect $\Delta_{\ell+1}\le\Delta_\ell$ for every $\ell\ge1$.
\end{prop}

At every full-memory evaluation time $k\ge\ell$, the memory-dependent upper bound in~\eqref{eq:memory_global_value_function_width} decreases with $\ell$ whenever $\Delta_\ell$ does, even vanishing for $\Delta_\ell\to0$ when the remaining horizon $H-k$ remains fixed. Yet, the above definition of $\Delta_\ell$ does not immediately enable its computation, nor offer a useful analytical characterization of its dependence on $\ell$. Section~\ref{sec:linear_gaussian_absorbing} offers an analytical, interpretable bound on $\Delta_\ell$ for linear-Gaussian dynamics.

\subsection{Linear-Gaussian systems}\label{sec:linear_gaussian_absorbing}\label{sec:compact_absorbing_system}

To apply the preceding bounds to the unbounded linear-Gaussian processes, we use compact state-space truncation, with the absorbing overflow-state construction of~\cite[Section~3.1]{Abate2010}.

Let the original Markov process evolve on $\mathbb{R}^n$ with transition density $t(\,\cdot\,|x)$ and initial measure $\pi_0$. Assume that the cost bounds $c_{\min},c_{\max}$ of Section~\ref{sec:filter_forgetting_value_function_widths} are finite and hold on all of $\mathbb{R}^n$. Following the cited construction, introduce an isolated absorbing state $\phi\notin\mathcal{X}$ and write $\mathcal{X}_{\phi}:=\mathcal{X}\cup\{\phi\}$. The resulting space is compact. Denote by $\tau_{\phi}$ the kernel that preserves transitions within $\mathcal{X}$, assigns all exit probability to $\phi$, and has $\tau_{\phi}(\cdot|\phi)=\delta_{\phi}$. The initial distribution likewise agrees with $\pi_0$ on $\mathcal{X}$ and assigns all initial mass outside $\mathcal{X}$ to $\phi$.

Set $c_{\phi}=c$ on $\mathcal{X}$ and $c_{\phi}(\phi)=c_{\min}$, and use the augmented refined partition $\mathcal{R}_{\phi}:=\mathcal{R}\cup\{\{\phi\}\}$. For memory length $\ell$, identify every memory state containing $\phi$ with the single absorbing overflow state $\phi$, giving $S_{\phi}^{(\ell)}:=\mathcal{R}^{\ell+1}\cup\{\phi\}$. By~\eqref{eq:robust_bellman_recursion}, its exact stage cost and deterministic self-loop give $\underline{V}_k(\phi)=\overline{V}_k(\phi)=(H-k)c_{\min}$, hence zero width. The interval constructions and soundness results of Sections~\ref{sec:concretesystemandabstractionproblem} and~\ref{sec:memory_dependent_imcs} apply to $(\tau_{\phi},c_{\phi})$ on $\mathcal{R}_{\phi}$. Let $\mathcal{I}_{\phi}$ denote a sound IMC abstraction obtained from this absorbing process. 

For the bounded stage costs considered here, agreement of the original and absorbing processes up to the first exit yields the following cost bound.

\begin{prop}[State-space truncation error]\label{prop:absorbing_exit_approximation}
	Let $J$ and $J_{\phi}$ denote the expected cumulative costs of the original and absorbing processes over the horizon $H$, and let $\tau_{\mathcal{X}}:=\inf\{j\ge0:x_j\notin\mathcal{X}\}$. Since the two processes agree before $\tau_{\mathcal{X}}$ and $0\le c(x)-c_{\min}\le C_c$, we have
	\begin{equation}
		0\le J-J_{\phi}\le C_c\sum_{j=0}^{H-1}\mathbb{P}(\tau_{\mathcal{X}}\le j).
		\label{eq:absorbing_exit_approximation}
	\end{equation}
	Consequently, sound bounds for the absorbing system imply the following corrected bounds for the original system
	\begin{equation}
		\underline{J}(\mathcal{I}_{\phi})\le J \le\overline{J}(\mathcal{I}_{\phi})+C_c\sum_{j=0}^{H-1}\mathbb{P}(\tau_{\mathcal{X}}\le j).
	\label{eq:absorbing_unbounded_performance_bounds}
	\end{equation}
\end{prop}

Any verified upper bounds on the exit probabilities may be substituted into these inequalities.

\subsubsection{Linear-Gaussian width bounds}\label{sec:linear_gaussian_specialized_width_bounds}

We now derive an explicit, fully computable robust value-function width bound for linear-Gaussian dynamics.

Consider
\begin{equation}
	x_{k+1}=Ax_k+w_k,\qquad w_k\sim\mathcal{N}(\hat{w},\Sigma_w),
	\label{eq:absorbing_linear_gaussian_system}
\end{equation}
with $\Sigma_w\succ0$ and absorption upon leaving $\mathcal{X}$, so that $t(x'|x)=\varphi_n(x';Ax+\hat{w},\Sigma_w)$, where $\varphi_n(\,\cdot\,;\bar{x},\Sigma)$ denotes the $n$-dimensional Gaussian density with mean $\bar{x}$ and covariance $\Sigma$.

For these dynamics, define the coefficient $\kappa$ as
\begin{equation}\label{eq:kappa_definition}
	\kappa:=\min\Bigl\{\|\Sigma_w\|_2^{1/2},\frac{d_M}{2}\Bigr\}\cdot\min\Bigl\{\bigl\|\Sigma_w^{-1/2}A\bigr\|_2,\frac{d_M}{2}\bigl\|\Sigma_w^{-1}A\bigr\|_2\Bigr\}.
\end{equation}

\begin{prop}[Linear-Gaussian robust value-function width bound]\label{prop:absorbing_linear_gaussian}
	Under~\eqref{eq:absorbing_linear_gaussian_system}, with $c$ $L_c$-Lipschitz, for every $k\in\{\ell,\ldots,H\}$,
	\begin{equation}\label{eq:gaussian_value_function_width_bound}
		D_k^{(\ell)}\le(H-k)\,d_M\min\{1,\kappa^\ell\}\left(L_c+\frac{G}{4}(H-k-1)C_c\right),
	\end{equation}
	with $\kappa$ as in~\eqref{eq:kappa_definition} and $G$ bounded by
	\begin{equation}
		G\le2^n(\sqrt{2/e})\left\|\Sigma_w^{-1/2}A\right\|_2 e^{\frac{d_M^2}{16}\left\|\Sigma_w^{-1/2}A\right\|_2^2}.
		\label{eq:gaussian_density_sensitivity_bound}
	\end{equation}
	If $\Sigma_w=\sigma^2I_n$, then $\kappa=\|A\|_2\min\{1,d_M^2/(4\sigma^2)\}$.
	%Every constant is computable directly from $(A,\Sigma_w,d_M,L_c,C_c,H,\ell)$.
\end{prop}

Proposition~\ref{prop:absorbing_linear_gaussian} identifies $\kappa$ as the contraction factor governing the dependence of the memory-dependent bound on the memory length $\ell$, i.e., how memory lengthening tightens the abstraction. In the isotropic case $\kappa$ is small when the one-step gain $\|A\|_2$ is small --- dynamics that strongly stabilize the departure state --- or when the grid is fine relative to the noise level, $d_M\ll\sigma$, since stochasticity alone then already blurs most of the dependence on the exact departure point. Conversely, when $A$ is expansive and the grid is coarse relative to the noise ($d_M\ge2\sigma$), $\kappa=\|A\|_2$ can reach or exceed one, and~\eqref{eq:gaussian_dispersion_bound} degenerates to the trivial bound $\Delta_\ell\le d_M$.

We do note that~\eqref{eq:gaussian_value_function_width_bound} is an upper bound, so a small $\kappa$ certifies that memory lengthening tightens the abstraction, whereas a large $\kappa$ leaves the bound uninformative without establishing that memory lengthening fails to tighten it. Moreove, even the relative sizes of~\eqref{eq:partition_global_value_function_width} and~\eqref{eq:gaussian_value_function_width_bound} do not directly order the exact widths. A small $\kappa$ therefore indicates when memory lengthening is certified to tighten the abstraction, rather than when it is preferable to partition refinement.

% \begin{rem}
% 	Memory is most valuable when $\kappa$ is small, so that $\Delta_\ell$ decays quickly with $\ell$. In the isotropic case this happens when the one-step gain $\|A\|_2$ is small --- dynamics that damp the departure state strongly --- or when the grid is fine relative to the noise level, $d_M\ll\sigma$, since stochasticity alone then already blurs most of the dependence on the exact departure point. Conversely, when $A$ is expansive and the grid is coarse relative to the noise ($d_M\ge2\sigma$), $\kappa=\|A\|_2$ can reach or exceed one, and~\eqref{eq:gaussian_dispersion_bound} degenerates to the trivial bound $\Delta_\ell\le d_M$: additional memory then no longer tightens the abstraction through this mechanism, and partition refinement is likely the more effective knob. This suggests using $\kappa$ as a practical guide for choosing between memory and refinement on a given system.
% \end{rem}

\section{Numerical examples}\label{sec:numerical_examples}

We evaluate the classic IMC abstraction, no-memory IMC abstraction, memory-dependent IMC abstraction, and partition-refined classic IMC abstraction on both (i) a variation of the linear-Gaussian benchmark used by Banse et al.~\cite{Banse2025Transfer}, and (ii) the example system described in Example~\ref{ex:running-example}. The abstractions considered are compared under different numbers of stored transition probability intervals, below referred to as simply \emph{(dynamic programming) complexity}.

\subsection{Linear-Gaussian benchmark}\label{sec:linear_gaussian_benchmark}

We introduce a second example, adapted from \cite{Banse2025Transfer}.
\begin{exmp}
	Consider the two-dimensional linear-Gaussian system $x_{k+1}=Ax_k+w_k$, $w_k\sim\mathcal{N}(0,\Sigma_w)$, with $x_0\sim\mathcal{N}(m_0,\Sigma_0)$, which we subject to
	\begin{equation*}
		A= \begin{bmatrix}
			0.117 & 0.005\\
			0 & 0.102
		\end{bmatrix}, \quad	m_0=\begin{bmatrix}
			0.32\\
			0.32
		\end{bmatrix},
	\end{equation*}
	at $\Sigma_0=0.01I_2$, $\Sigma_w=0.174I_2$, and a horizon of $H=16$. On $\mathcal{X}$, the stage cost is set as	$c(x)=\min\{x^\top x,4\}$.
\end{exmp}

The compact truncation~\cite{Abate2010} is executed with $\mathcal{X}=[-6,6]^2$. Each coordinate axis is divided into $M=9$ intervals, yielding $81$ basis cells. In the implementation, the correction at horizon $k$ is applied using $\varepsilon_{\mathcal{X},k}:=C_c\sum_{j=0}^{k-1}\widehat{p}_j$, where $\widehat{p}_j$ is an upper bound on $\mathbb{P}(\tau_{\mathcal{X}}\le j)$ obtained by summing the exact one-dimensional Gaussian marginal tail probabilities over both coordinates and all times up to $j$. Write $\underline{J}_k(\mathcal{I}_{\phi})$ and $\overline{J}_k(\mathcal{I}_{\phi})$ for the robust expected-cost bounds obtained from the same absorbing-system IMC at horizon $k$. Thus, every displayed interval $[\underline{J}_k(\mathcal{I}_{\phi}),\overline{J}_k(\mathcal{I}_{\phi})+\varepsilon_{\mathcal{X},k}]$ encloses the expected cost of the original unbounded process. As the system is fixed, the same correction is added to every upper bound, so differences between the abstractions result exclusively from their stage-cost and transition-probability interval differences.

\begin{figure}[tb]
	\centering
\includegraphics[width=0.57\columnwidth]{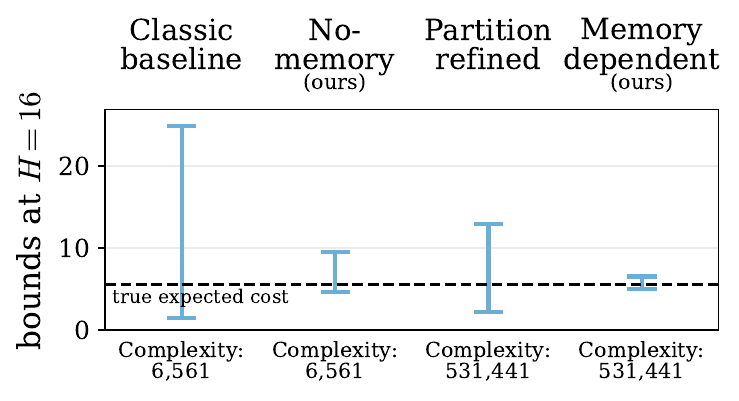}
	\caption{Matched-complexity expected-cost intervals for the modified linear-Gaussian system at $H=16$. The classic IMC and no-memory member store $81^2=6,\!561$ probability intervals between non-absorbing states; partition refinement and the memory-dependent IMC with memory length one store $729^2=81^3=531,\!441$. Every non-absorbing transition probability distribution also contains one exit interval.}
	\label{fig:banse_variation_bound_gap_bars}
\end{figure}

For the matched comparison at $M=9$, the classic IMC uses $N=1$ and memory length $\ell=0$, partition refinement uses $N=3$ and gives $729$ refined current cells, and the no-memory and memory-dependent members both use $N=1$ and memory length $\ell=1$, with $81$ current cells. The partition-refined and memory-dependent constructions therefore use an equal number of stored non-absorbing transition probability intervals, namely $729^2=81^3=531441$. Each non-absorbing transition probability vector additionally contains one exit interval, while the absorbing overflow state has an exact self-loop. Figure~\ref{fig:banse_variation_bound_gap_bars} reports the expected-cost intervals at $H=16$. The partition-refined and memory-dependent intervals have widths $10.6731$ and $1.5295$, respectively. At this matched complexity budget, the memory-dependent approach therefore yields a substantially tighter interval than the partition-refined abstraction.

\subsection{Simple example system}\label{sec:example_system_experiment}

\begin{figure}[tb]
	\centering
\includegraphics[width=0.57\columnwidth]{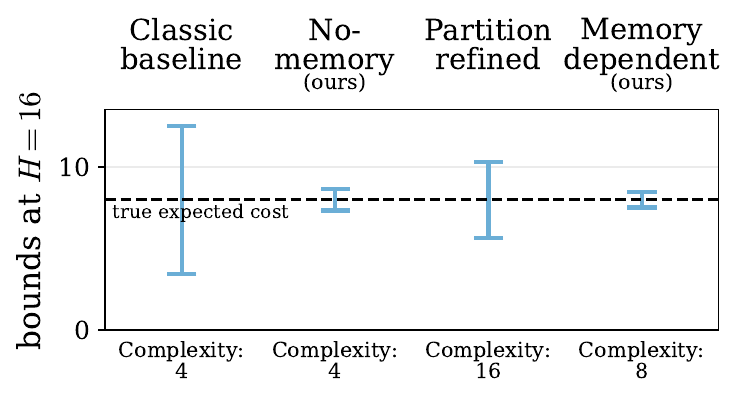}
	\caption{Expected-cost intervals for system~\eqref{eq:example_system} at $H=16$. Complexity counts stored transition probability intervals. The dashed line marks the concrete expected cost $J=8$.}
	\label{fig:example_bound_gap_bars}
\end{figure}

For Example~\ref{ex:running-example} at $M=2$, we compare the baseline $(N,\ell)=(1,0)$ (see Fig.~\ref{fig:basis_system}), partition refinement $(N,\ell)=(2,0)$ (see Fig.~\ref{fig:partition_refinement}), and the memory-dependent (see Fig.~\ref{fig:memory_dependent_abstraction}) and no-memory constructions at $(N,\ell)=(1,1)$. We evaluate the abstractions at $H=16$. Due to the simplicity of the system equations, we are able to apply~\eqref{eq:robust_bellman_recursion} to the computed intervals using exact rational arithmetic, i.e., without requiring numeric methods. Figure~\ref{fig:example_bound_gap_bars} report the resulting bounds. The memory-dependent abstraction is now tightest, followed by no-memory, partition refinement, and the baseline. The comparison uses $16$ stored transition intervals for partition refinement, $8$ shift-compatible intervals for memory, and $4$ intervals each for the baseline and no-memory constructions, clearly demonstrating that even for this simple system, the memory-dependent abstraction can outperform partition refinement at a matched complexity budget. We note that predicting a-priori which approach will yield tighter bounds remains unresolved and is a compelling direction for future research.

\section{Conclusions and future work}\label{sec:conclusion}

We develop memory-dependent IMC abstractions that provide sound bounds on finite-horizon expected cumulative costs for continuous-state stochastic systems. Conditioning on previously visited cells restricts the possible current-state distributions, tightening local cost and transition intervals. A second, no-memory construction, exploits transition-kernel information in a similar way, but does not enlarge the abstract state space. Under Lipschitz assumptions, we derive separate robust value-function width bounds governed by the refined-cell diameter and the remaining distributional ambiguity $\Delta_\ell$. For linear-Gaussian dynamics, the analytically computable coefficient $\kappa$ guarantees geometric decay of the memory-dependent bound as $\ell$ increases along full-memory stages with fixed remaining horizon when $\kappa<1$. Numerical examples show substantially tighter expected-cost widths for memory than for partition refinement at equal transition-storage complexity on the linear-Gaussian benchmark, and substantially tighter bounds with fewer stored intervals for memory on the simple example system. These results establish memory as an alternative to partition refinement, while a general ordering of their expected-cost widths remains unresolved.

A central direction for future work is to identify which properties of the concrete system govern the relative effectiveness of memory lengthening and partition refinement in tightening expected-cost bounds. The linear-Gaussian analysis relates the derived bounds to the dynamics and noise level, but these upper bounds do not order the exact expected-cost interval widths. Establishing criteria for this comparison at a given complexity budget could guide the choice between additional memory and finer partitions.

The currently presented canonical form assumes a uniform partitioning over all dimensions of the state space. In practice, this is rarely the optimal partition of concrete states into the finite number of bins available. The simplification is employed, as a systematic binning optimization approach is yet unavailable in literature. Nonetheless, this aspect of the problem remains highly relevant, and cannot be omitted from a definitive treatment of the two methods. Stronger still, positive memory lengths cause partition decisions to additionally affect previous-state information, making bin optimization even more relevant to the resulting accuracy.

\appendix
\section{Technical results and proofs}\label{sec:proofs}

\begin{pf*}{Proof of Proposition~\ref{prop:recursivity}}
	Induction on $j=0,1,\dots,\ell$, from $\beta_{-\ell}^{\xi,m}=\delta_\xi$, yields
	\begin{equation*}
	\begin{aligned}
		\beta_{-\ell+j}^{\xi,m}(A)=\mathbb{P}\bigl(x_{-\ell+j}&\in A,\\
		x_{-\ell+i}&\in r_{-\ell+i}\text{ for }i=1,\ldots,j |x_{-\ell}=\xi\bigr).
	\end{aligned}
	\end{equation*}
	At $j=\ell$ for $A=r_0$, we then have exactly $\beta_0^{\xi,m}(r_0)=\mathbb{P}(E_m|x_{-\ell}=\xi)$; the normalization constant for the conditional measure of $x_0$~\eqref{eq:normalization_at_the_end}. Finally, a single extra application of the transition kernel $\tau$ gives
	\begin{equation*}
		\mathbb{P}(x_{+1}\in r'|E_m,x_{-\ell}=\xi)=\int_{r_0}\tau(r'|x_0)\,\mathbb{P}(\mathrm{d}x_0|E_m,x_{-\ell}=\xi),
	\end{equation*}
		and the final identity can be proven by substitution of the normalized measure.
\end{pf*}

\begin{pf*}{Proof of Theorem~\ref{thm:memory_dependent_soundness}}
	While the concrete $\mathcal{M}_{\mathcal{X}}$-process $(x_k)_k$ is Markov, its finite-memory observation $(m_k)_k$ generally is not. Yet we construct next a \emph{time-varying} finite-state MC $\mathcal{M}^{(\ell)}$, with state process $(\widehat{m}_k)_k$, whose time-$k$ stage costs and transition rows equal the conditional expected costs and one-step conditional probabilities induced by $(m_k)_k$ (or equivalently, $(x_k)_k$).

	We construct $\mathcal{M}^{(\ell)}$ such that $m_k$ (the discrete observations of the concrete process $\mathcal{M}_{\mathcal{X}}$) and $\widehat{m}_k$ (its discrete-space time-varying MC clone) have the same distribution at every time $k$. We then show that the time-varying transition probabilities and stage costs of $\mathcal{M}^{(\ell)}$ are admissible, i.e., $\mathcal{M}^{(\ell)}\in\mathcal{I}^{(\ell)}$, under the memory-dependent IMC abstraction from Definition~\ref{def:memory_dependent_imc_abstraction}. Together, these properties yield
	\begin{equation*}
		\mathcal{M}^{(\ell)}\in\mathcal{I}^{(\ell)}\qquad\text{and}\qquad J(\mathcal{M}^{(\ell)})=J(\mathcal{M}_{\mathcal{X}}).
	\end{equation*}

	Let $r_k\in\mathcal{R}$ denote the cell containing $x_k$, and let $m_k$ denote the corresponding memory state generated by the concrete process. We take the first $\ell$ entries of the initial memory state $m_0$ to be auxiliary and uniformly distributed independently of $x_0$, as encoded by the initial distribution $\alpha^{(\ell)}$ in Definition~\ref{def:memory_dependent_imc_abstraction}. Thus, we initialize $\widehat{m}_0$ with the same distribution as $m_0$, as $\mathbb{P}(\widehat{m}_0=m):=\mathbb{P}(m_0=m)=\alpha^{(\ell)}(m)$. For every memory state $m\in\mathcal{R}^{\ell+1}$ reachable at time $k$, define
	\begin{equation*}
	\begin{split}
		c_k^{(\ell)}(m)&:=\mathbb{E}[c(x_k)|m_k=m], \\
		P_k^{(\ell)}(m,m')&:=\mathbb{P}(m_{k+1}=m'|m_k=m),
	\end{split}
	\end{equation*}
	for all $m'\in\mathcal{R}^{\ell+1}$. At unreachable memories, fix arbitrary admissible entries; these exist by the definition of an IMC but do not affect the state marginals or expected cost.

	Together with the initial distribution specified above, the stage costs $c_k^{(\ell)}$ and transition rows $P_k^{(\ell)}$ define the time-varying MC $\mathcal{M}^{(\ell)}$. Its state marginals agree with those of $m_k$ by construction. Consequently, the Markov property allows us to decompose the expected cost as
	\begin{equation*}
	\begin{aligned}
		J(\mathcal{M}^{(\ell)})&=\sum_{k=0}^{H-1}\sum_m\mathbb{P}(\widehat{m}_k=m)c_k^{(\ell)}(m)\\
		&=\sum_{k=0}^{H-1}\sum_m\mathbb{P}(m_k=m)\mathbb{E}[c(x_k)|m_k=m]\\
		&=\sum_{k=0}^{H-1}\mathbb{E}[c(x_k)]=J(\mathcal{M}_{\mathcal{X}}),
	\end{aligned}
	\end{equation*}
	where the second equality holds by construction.

	It remains to show that the stage costs and transition probabilities of $\mathcal{M}^{(\ell)}$ are admissible under $\mathcal{I}^{(\ell)}$. Fix $k\in\{\ell,\ldots,H-1\}$ and a reachable memory state $m$, and let $\mu^{m}_{k}$ denote the conditional measure over the oldest concrete state $x_{k-\ell}$, defined by
	\begin{equation*}
	\mu^{m}_{k}(A):=\mathbb{P}(x_{k-\ell}\in A|m_k=m),
	\end{equation*}
	for every measurable $A\subseteq\mathcal{X}$.

	Conditioning on $x_{k-\ell}$ and invoking the Markov property expresses the following quantities as weighted averages with respect to $\mu_k^m$. Their components lie in the corresponding abstraction intervals by Definition~\ref{def:memory_dependent_imc_abstraction}, and convexity of these intervals therefore gives
	\begin{equation*}
		\underline{c}_{k}(m)\hspace{-1mm}\le\hspace{-1mm}\int_{r_{k-\ell}}\hspace{-1.5mm}\left[\int_{r_k}\hspace{-2mm}c(x)\,\mu_{\xi,m}^{(\ell)}(\mathrm{d}x)\right]\hspace{-0.8mm}\mu^{m}_{k}(\mathrm{d}\xi)\le\overline{c}_{k}(m),
	\end{equation*}
	where the middle expression equals $c_k^{(\ell)}(m)$.

	Similarly, for every shift-compatible $m'$, with $r'$ denoting the cell appended to form $m'$,
	\begin{equation*}
		\underline{P}(m,m')\le\int_{r_{k-\ell}}\left[\int_{r_k}\tau(r'|x)\,\mu_{\xi,m}^{(\ell)}(\mathrm{d}x)\right]\mu^{m}_{k}(\mathrm{d}\xi)\le\overline{P}(m,m'),
	\end{equation*}
	where the middle expression equals $P_k^{(\ell)}(m,m')$. If $m'$ is not shift-compatible with $m$, then $P_k^{(\ell)}(m,m')=0=\underline{P}(m,m')=\overline{P}(m,m')$.

	For $0\leq k<\min\{\ell,H\}$, let $D_k\in\mathcal R^{\ell-k}$ be a (uniformly selected) dummy prefix, independent of the concrete process. The full memory state is then $m_k=(D_k,\sigma_k(m))$. The prefix value is arbitrary from the perspective of the observed process, which has generated no information beyond the available suffix $\sigma_k(m)=(r_0,\ldots,r_k)$. All full states $m\in\mathcal{R}^{\ell+1}$ sharing $\sigma_k(m)$ therefore represent the same available information and receive the same cost interval. Let $P(\sigma_k(m),r'_{k+1})$ denote the concrete transition probability of appending $r'_{k+1}$, conditioned on this suffix. Each of the $|\mathcal R|^{\ell-k-1}$ possible dummy prefixes $D_{k+1}\in\mathcal R^{\ell-k-1}$ determines a unique, yet equivalently behaving successor $m'=(D_{k+1},\sigma_k(m),r'_{k+1})$. By distributing the suffix transition probability uniformly over these successors, as $P_k^{(\ell)}(m,m'):=\frac{P(\sigma_k(m),r'_{k+1})}{|\mathcal{R}|^{\ell-k-1}}$, the preceding mixture argument places the suffix-conditioned cost and transition probability in their intervals for memory length $k$. Aggregating these divided intervals over all $|\mathcal{R}|^{\ell-k-1}$ dummies recovers the original transition interval for memory length $k$, so the preceding mixture and admissibility arguments apply unchanged.

	Picking arbitrary admissible entries at unreachable memories, the interval inclusions established above imply $\mathcal{M}^{(\ell)}\in\mathcal{I}^{(\ell)}$. Together with the cost equality established above, Equation~\eqref{eq:imc_member_cost_extrema} now yields the soundness claim~\eqref{eq:memory_dependent_soundness}.
\end{pf*}

\begin{pf*}{Proof of Lemma~\ref{lem:memory_cellwise_interval_inclusion}}
	Fix $\ell$, time $k$, $m_k=(r_{k-\ell},\ldots,r_k)$, and a shift-compatible $m_{k+1}=(r_{k-\ell+1},\ldots,r_k,r_{k+1})$. First suppose that $m_k$ is reachable. For every admissible $\xi\in r_{k-\ell}$, the measure $\mu_{\xi,m_k}^{(\ell)}$ is supported on $r_k$. Indeed, for $\ell\ge1$ this follows because $E_{m_k}$ includes the observation $x_k\in r_k$, whereas for $\ell=0$ the event is vacuous and $x_k=\xi\in r_k$. Hence, the refined-cell definitions and the Markov property give
	\begin{equation*}
	\underline{c}_{k}(r_k)\le\int_{r_k}c(x)\,\mu_{\xi,m_k}^{(\ell)}(\mathrm{d}x)\le\overline{c}_{k}(r_k),
	\end{equation*}
	and
	\begin{equation*}
		\underline{P}(r_k,r_{k+1})
		\le\int_{r_k}\tau(r_{k+1}|x)\,\mu_{\xi,m_k}^{(\ell)}(\mathrm{d}x)\le \overline{P}(r_k,r_{k+1}).
	\end{equation*}
	Taking the infimum and supremum over the same nonempty set of admissible $\xi$ yields~\eqref{eq:memory_cost_inclusion} and~\eqref{eq:memory_transition_inclusion}. If $m_k$ is unreachable, the prescribed fallback intervals equal the corresponding refined-cell intervals, so both inclusions hold with equality at their outer endpoints.
\end{pf*}

\begin{pf*}{Proof of Theorem~\ref{thm:absorbing_value_function_width_convergence}}
	We bound the local cost and transition widths and propagate them through the robust Bellman recursion.

	\emph{Local cost width.} Fix a full-memory stage $k\in\{\ell,\ldots,H-1\}$ and $m=(y_{-\ell},\ldots,y_0)\in\mathcal{Y}^{\ell+1}$. By Kantorovich--Rubinstein duality~\cite{KantorovichRubinstein1958}, the Lipschitz property of $c$ gives, for every admissible $\xi,\zeta\in y_{-\ell}$,
	\begin{equation*}
		\left|\int c\,\mathrm{d}\mu_{\xi,m}^{(\ell)}-\int c\,\mathrm{d}\mu_{\zeta,m}^{(\ell)}\right|\le L_cW_1(\mu_{\xi,m}^{(\ell)},\mu_{\zeta,m}^{(\ell)})\le L_c\Delta_\ell,
	\end{equation*}
	where the last step is Definition~\ref{def:memory_conditioned_dispersion}. Taking the supremum over $\xi$ and the infimum over $\zeta$ gives
	\begin{equation}
		\overline{c}_{k}(m)-\underline{c}_{k}(m)\le L_c\Delta_\ell.
		\label{eq:memory_cost_width}
	\end{equation}

	\emph{Local transition width.} For every $y\in\mathcal{Y}$, Assumption~\ref{ass:cellwise_transition_density_sensitivity} gives
	\begin{equation*}
		\int_{\mathcal{X}}\left|t\left(x'|x\right)-t\left(x'|\tilde{x}\right)\right|\,\mathrm{d}x' \leq G_y\|x-\tilde{x}\|_2,
	\end{equation*}
	for all $x,\tilde{x}\in y$. For $y,y'\in\mathcal{Y}$, set
	\begin{equation*}
		G_{yy'}:=\int_{y'}g_y(x')\,\mathrm{d}x'.
	\end{equation*}
	Assumption~\ref{ass:cellwise_transition_density_sensitivity} gives
	\begin{equation*}
		|\tau(y'|x)-\tau(y'|\tilde{x})|\le G_{yy'}\|x-\tilde{x}\|_2,
	\end{equation*}
	for all $x,\tilde{x}\in y$. For the fixed memory state $m$, let $m'=(y_{-\ell+1},\ldots,y_0,y')$ denote the shift-compatible successor associated with $y'\in\mathcal{Y}$. Consequently,
	\begin{equation}
		\overline{P}(m,m')-\underline{P}(m,m')\le G_{y_0y'}\sup_{\xi,\zeta}W_1(\mu_{\xi,m}^{(\ell)},\mu_{\zeta,m}^{(\ell)})\le G_{y_0y'}\Delta_\ell.
	\end{equation}
	Here, the supremum is over admissible $\xi,\zeta\in y_{-\ell}$, and the second inequality is Definition~\ref{def:memory_conditioned_dispersion}.
	Summing over $y'\in\mathcal{Y}$ and using
	\begin{equation}
		\sum_{y'\in\mathcal{Y}}G_{y_0y'}=\int_{\mathcal{X}}g_{y_0}(x')\,\mathrm{d}x'=G_{y_0}\le G,
	\end{equation}
	yields
	\begin{equation}
		\sum_{y'\in\mathcal{Y}}\left(\overline{P}(m,m')-\underline{P}(m,m')\right)\le G\Delta_\ell.
		\label{eq:memory_transition_width}
	\end{equation}

	\emph{Propagation through the Bellman recursion.} Equations~\eqref{eq:memory_cost_width} and~\eqref{eq:memory_transition_width} bound the memory-dependent local cost width by $L_c\Delta_\ell$ and the sum of the componentwise transition widths by $G\Delta_\ell$. For a partition-refined cell of diameter $d_M/N$, the same Lipschitz arguments applied to the family of point masses supported on that cell give the corresponding bounds $L_cd_M/N$ and $Gd_M/N$.

	At time $k$, let $[\underline{p}_{y'},\overline{p}_{y'}]$ be the components of an admissible abstract transition interval. For any two feasible distributions $p^{(1)}$ and $p^{(2)}$,
	\begin{equation}
	\|p^{(1)}-p^{(2)}\|_{\mathrm{TV}}:=\frac{1}{2}\sum_{y'}|p_{y'}^{(1)}-p_{y'}^{(2)}|\le\frac{1}{2}\sum_{y'}(\overline{p}_{y'}-\underline{p}_{y'}).
		\label{eq:transition_vector_tv}
	\end{equation}
	Let $p^+$ and $p^-$ be the maximizing and minimizing transition distributions in the corresponding robust Bellman updates, and write $(p)^{\top}V:=\sum_{y'} p_{y'}V(y')$. Every time-$(k+1)$ cost-to-go value lies in $[(H-k-1)c_{\min},(H-k-1)c_{\max}]$. Thus, for $\star\in\{(\ell),\mathrm{pr}\}$,
	\begin{equation}
	\begin{split}
		&(p^+)^{\top}\overline{V}_{k+1}^{\star}-(p^-)^{\top}\underline{V}_{k+1}^{\star}\\
		&\hspace{14mm}=(p^+)^{\top}\left(\overline{V}_{k+1}^{\star}-\underline{V}_{k+1}^{\star}\right)+\left(p^+-p^-\right)^{\top}\underline{V}_{k+1}^{\star}\\
		&\hspace{14mm}\le D_{k+1}^{\star}+\|p^+-p^-\|_{\mathrm{TV}}(H-k-1)C_c.
	\end{split}
		\label{eq:bellman_width_decomposition}
	\end{equation}
	Combining~\eqref{eq:bellman_width_decomposition} with~\eqref{eq:transition_vector_tv} and the local cost widths gives
	\begin{equation}
	\begin{split}
		D_k^{(\ell)}&\le D_{k+1}^{(\ell)}+\Delta_\ell\left(L_c+\frac{G}{2}(H-k-1)C_c\right),
	\end{split}
		\label{eq:value_function_width_recursions_ell}
	\end{equation}
	\begin{equation}
	\begin{split}
		D_k^{\mathrm{pr}}&\le D_{k+1}^{\mathrm{pr}}+\frac{d_M}{N}\left(L_c+\frac{G}{2}(H-k-1)C_c\right).
	\end{split}
		\label{eq:value_function_width_recursions_pr}
	\end{equation}
	Since $D_H^{(\ell)}=D_H^{\mathrm{pr}}=0$ and
	\begin{equation}
		\sum_{j=k}^{H-1}(H-j-1)=\frac{(H-k)(H-k-1)}{2},
	\end{equation}
	summing~\eqref{eq:value_function_width_recursions_ell} proves~\eqref{eq:memory_global_value_function_width}, and~\eqref{eq:value_function_width_recursions_pr} proves~\eqref{eq:partition_global_value_function_width}.
\end{pf*}

\begin{pf*}{Proof of Proposition~\ref{prop:dispersion_monotone}}
	Fix $\ell\ge1$, $m' = (y_{-\ell-1},\ldots,y_0)\in\mathcal{Y}^{\ell+2}$, and admissible $\xi',\zeta'\in y_{-\ell-1}$, and write $m:=(y_{-\ell},\ldots,y_0)$ for the length-$(\ell+1)$ suffix of $m'$, so that $E_{m'}=\{x_{-\ell}\in y_{-\ell}\}\cap E_m$. Conditioning on $x_{-\ell}$ and using the Markov property gives, for every measurable $A\subseteq y_0$,
	\begin{equation*}
	\begin{split}
		\mu_{\xi',m'}^{(\ell+1)}(A)&=\int_{y_{-\ell}}\mu_{\xi,m}^{(\ell)}(A)\,\pi_{\xi'}(\mathrm{d}\xi),\\
		\pi_{\xi'}(\mathrm{d}\xi)&:=\frac{\mathbb{P}(E_m|x_{-\ell}=\xi)\,\tau(\mathrm{d}\xi|\xi')}{\mathbb{P}(E_{m'}|x_{-\ell-1}=\xi')},
	\end{split}
	\end{equation*}
	a probability measure on $y_{-\ell}$, well defined since $\xi'$ is admissible; likewise $\mu_{\zeta',m'}^{(\ell+1)}=\int_{y_{-\ell}}\mu_{\zeta,m}^{(\ell)}\,\pi_{\zeta'}(\mathrm{d}\zeta)$. For $\xi,\zeta\in y_{-\ell}$, let $\gamma_{\xi,\zeta}\in\Gamma\bigl(\mu_{\xi,m}^{(\ell)},\mu_{\zeta,m}^{(\ell)}\bigr)$ be a coupling attaining the infimum in the definition~\eqref{eq:wasserstein_distance} of $W_1$, that is, $\int\|x-\tilde{x}\|_2\,\gamma_{\xi,\zeta}(\mathrm{d}x,\mathrm{d}\tilde{x})=W_1\bigl(\mu_{\xi,m}^{(\ell)},\mu_{\zeta,m}^{(\ell)}\bigr)$, chosen measurably in $(\xi,\zeta)$; such a minimizer exists as both measures are supported on the compact cell $y_0$ and the cost $\|x-\tilde{x}\|_2$ is continuous. Then $\int\int\gamma_{\xi,\zeta}\,\pi_{\xi'}(\mathrm{d}\xi)\pi_{\zeta'}(\mathrm{d}\zeta)$ is a coupling of $\mu_{\xi',m'}^{(\ell+1)}$ and $\mu_{\zeta',m'}^{(\ell+1)}$, so
	\begin{equation*}
		W_1\bigl(\mu_{\xi',m'}^{(\ell+1)},\mu_{\zeta',m'}^{(\ell+1)}\bigr)\le\int_{y_{-\ell}}\int_{y_{-\ell}}W_1\bigl(\mu_{\xi,m}^{(\ell)},\mu_{\zeta,m}^{(\ell)}\bigr)\pi_{\xi'}(\mathrm{d}\xi)\,\pi_{\zeta'}(\mathrm{d}\zeta)\ \le\ \Delta_\ell.
	\end{equation*}
	Taking the supremum over admissible $\xi',\zeta'\in y_{-\ell-1}$ and the maximum over $m'\in\mathcal{Y}^{\ell+2}$ gives $\Delta_{\ell+1}\le\Delta_\ell$.
\end{pf*}

\begin{pf*}{Proof of Proposition~\ref{prop:absorbing_exit_approximation}}
	The two trajectories and their accumulated costs agree before $\tau_{\mathcal{X}}$. At every later stage, the difference between the original-process cost and $c_{\min}$ lies in $[0,C_c]$. Hence, pathwise,
	\begin{equation*}
		0\le\sum_{j=0}^{H-1}c(x_j)-\sum_{j=0}^{H-1}c_{\phi}(x_j^\phi)\le C_c\sum_{j=0}^{H-1}\mathbf{1}_{\{\tau_{\mathcal{X}}\le j\}}.
	\end{equation*}
	Taking expectations proves~\eqref{eq:absorbing_exit_approximation}. Combining this estimate with $\underline{J}(\mathcal{I}_{\phi})\le J_{\phi}\le\overline{J}(\mathcal{I}_{\phi})$ gives~\eqref{eq:absorbing_unbounded_performance_bounds}.
\end{pf*}

\begin{pf*}{Proof of Proposition~\ref{prop:absorbing_linear_gaussian}}
	We verify Assumption~\ref{ass:cellwise_transition_density_sensitivity} and bound $G$ and $\Delta_\ell$ before applying Theorem~\ref{thm:absorbing_value_function_width_convergence}.

	\emph{Transition-density sensitivity.} The Gaussian transition density $t(x'|x)=\varphi_n(x';Ax+\hat{w},\Sigma_w)$ is strictly positive and continuous. The current-state gradient of the Gaussian transition density is
	\begin{equation}\label{eq:gaussian_transition_gradient}
		\nabla_x t(x'|x)=A^\top\Sigma_w^{-1}(x'-Ax-\hat{w})\varphi_n(x';Ax+\hat{w},\Sigma_w).
	\end{equation}
	Let $x_y^{\mathrm{c}}$ be the center of $y$, and let the noise-normalized displacement that the deviation of a current-state $x$ from the center $x_y^{\mathrm{c}}$ induces in the next-state conditional mean, be defined
	\[
		d_{x}:=\Sigma_w^{-1/2}(Ax+\hat{w}-(Ax_y^{\mathrm{c}}+\hat{w})) =\Sigma_w^{-1/2}A(x-x_y^{\mathrm{c}}).
	\]
	Its maximum over all $x\in y$ can be upper bounded as
	\begin{equation}
	\begin{split}
		\rho_y&:=\sup_{x\in y}\left\|d_{x}\right\|_2\le\frac{d_M}{2}\left\|\Sigma_w^{-1/2}A\right\|_2.
	\end{split}
	\end{equation}
	Let us further denote the noise-normalized displacement of a known next state $x'$ from the mean next state associated with the center $Ax_y^{\mathrm{c}}+\hat{w}$, as 
	\[
	d_{x'}:=\Sigma_w^{-1/2}(x'-(Ax_y^{\mathrm{c}}+\hat{w})).
	\]
	Equation~\eqref{eq:gaussian_transition_gradient} and the inequalities
	\begin{equation}
	\begin{aligned}
		\|d_{x'}-d_{x}\|_2 e^{-\frac{1}{2}\|d_{x'}-d_{x}\|_2^2}&\le(\sqrt{2/e})e^{-\frac{1}{4}\|d_{x'}-d_{x}\|_2^2},\\
		\|d_{x'}-d_{x}\|_2^2&\ge\tfrac12\|d_{x'}\|_2^2-\|d_{x}\|_2^2,
	\end{aligned}\label{eq:label_not_sure_what_to_call}
	\end{equation}
	give, uniformly over $x\in y$,
	\begin{equation*}
		\|\nabla_x t(x'|x)\|_2 \le (2\pi)^{-\frac{1}{2}n}\det(\Sigma_w)^{-\frac{1}{2}}\|\Sigma_w^{-1/2}A\|_2(\sqrt{2/e}) e^{\frac{1}{4}(\rho_y^2-\frac{1}{2}\|d_{x'}\|_2^2)}=:g_y(x').
	\end{equation*}
	In~\eqref{eq:label_not_sure_what_to_call}, the first scalar inequality follows by maximizing the left-hand side with respect to $\|d_{x'}-d_{x}\|_2$, and the second follows from $\|d_{x'}\|_2^2\le2\|d_{x'}-d_{x}\|_2^2+2\|d_{x}\|_2^2$. As $y$ is convex, the mean-value theorem gives $|t(x'|x)-t(x'|\tilde{x})|\le g_y(x')\|x-\tilde{x}\|_2$, proving~\eqref{eq:cellwise_transition_density_sensitivity}. Finally, changing variables from $x'$ to $d_{x'}$, with $\mathrm{d}x'=\det(\Sigma_w)^{1/2}\mathrm{d}d_{x'}$, yields
	\begin{equation}
		G_y=2^n(\sqrt{2/e})\|\Sigma_w^{-1/2}A\|_2 e^{\frac{1}{4}\rho_y^2}\le2^n(\sqrt{2/e})\|\Sigma_w^{-1/2}A\|_2 e^{\frac{1}{16}d_M^2\|\Sigma_w^{-1/2}A\|_2^2},
	\end{equation}
	where $(2\pi)^{-n/2}\int_{\mathbb{R}^n}e^{-\frac{1}{8}\|d_{x'}\|_2^2}\mathrm{d}d_{x'}=2^n$. Taking the maximum over $y$ proves~\eqref{eq:gaussian_density_sensitivity_bound}.

	\emph{Remaining distributional ambiguity.} We next show that, for every $\ell\ge1$ and every $m\in\mathcal{Y}^{\ell+1}$,
	\begin{equation}
	\begin{split}
		\Delta_\ell(m)&\le\kappa^\ell d_M,\\
		\Delta_\ell&\le d_M\min\{1,\kappa^\ell\},
	\end{split}
		\label{eq:gaussian_dispersion_bound}
	\end{equation}
	with $\kappa$ as in~\eqref{eq:kappa_definition}.

	\emph{Backward-tail functions and kernels.} For $m=(y_{-\ell},\ldots,y_0)\in\mathcal{Y}^{\ell+1}$, define the backward-tail functions for $j=-\ell,\ldots,-1$,
	\begin{equation}
		b(y_{j+1:0}|x_j):=\int_{y_{j+1}}\hspace{-3mm}b(y_{j+2:0}|x_{j+1})\,t(x_{j+1}|x_j)\,\mathrm{d}x_{j+1},
		\label{eq:memory_forward_tail}
	\end{equation}
	so that $b(y_{j+1:0}|x_j)$ is the probability of observing the tail $y_{j+1},\ldots,y_0$ of $m$ conditional on $x_j$, having $b(y_{0+1:0}|\,\cdot\,)\equiv1$. As $t$ is everywhere positive, $b(y_{j+1:0}|x_j)>0$ for every $j$ and every $x_j\in y_j$, so every $m\in\mathcal{Y}^{\ell+1}$ is admissible at every oldest state. For every $x_j\in y_j$ and measurable $A\subseteq y_{j+1}$, define the kernels
	\begin{equation*}
		K_j^m(A|x_j):=\frac{\int_A b(y_{j+2:0}|x_{j+1})\,t(x_{j+1}|x_j)\,\mathrm{d}x_{j+1}}{b(y_{j+1:0}|x_j)}.
	\end{equation*}

	\emph{Claim 1. Backward decomposition.} For $\xi\in y_{-\ell}$, $\mu_{\xi,m}^{(\ell)}=\delta_\xi K_{-\ell}^mK_{-\ell+1}^m\cdots K_{-1}^m$. Indeed, by the Markov property, $b(y_{j+1:0}|x_j)=\mathbb{P}(x_{j+1}\in y_{j+1},\ldots,x_0\in y_0|x_j)$, so for measurable $A\subseteq y_{j+1}$,
	\begin{equation*}
		\mathbb{P}\left(x_{j+1}\in A|x_j,\ x_{j+1}\in y_{j+1},\ldots,x_0\in y_0\right)=\frac{\int_A b(y_{j+2:0}|x_{j+1})\,t(x_{j+1}|x_j)\,\mathrm{d}x_{j+1}}{b(y_{j+1:0}|x_j)}=K_j^m(A|x_j).
	\end{equation*}
	Again by the Markov property, conditioning on $x_j$ makes the earlier observations irrelevant to the future, so under $\mathbb{P}(\,\cdot\,|E_m,x_{-\ell}=\xi)$ the sequence $(x_{-\ell},\ldots,x_0)$ is a Markov chain started at $\delta_\xi$ with transition kernels $K_{-\ell}^m,\ldots,K_{-1}^m$. As the terminal state of this conditioned Markov chain is $x_0$, its distribution is $\delta_\xi K_{-\ell}^mK_{-\ell+1}^m\cdots K_{-1}^m$. By the definition of $\mu_{\xi,m}^{(\ell)}$, this terminal distribution is exactly $\mu_{\xi,m}^{(\ell)}$, which proves the claim.

	\emph{Claim 2. Log-concavity of the backward functions.} Each $b(y_{j+2:0}|\,\cdot\,)$ is log-concave on $\mathbb{R}^n$ and bounded by $1$. Boundedness holds since $b$ is a probability. For $j=-1$ the function is identically $1$. For $j\le-2$, unrolling~\eqref{eq:memory_forward_tail} gives,
	\begin{equation*}
		b(y_{j+2:0}|x_{j+1})=\hspace{-2mm}\int\limits_{\mathbb{R}^{n(|j|-1)}}\hspace{-1mm}\prod_{i=j+1}^{-1}\hspace{-2mm}\varphi_n\bigl(x_{i+1};Ax_i+\hat{w},\Sigma_w\bigr)\mathbf{1}_{y_{i+1}}(x_{i+1})\,\mathrm{d}x_{j+2:0}.
	\end{equation*}
	Each Gaussian factor is the exponential of a concave quadratic form in the pair $(x_i,x_{i+1})$, hence log-concave jointly; each $\mathbf{1}_{y_{i+1}}$ is the indicator of a convex cell, hence log-concave. The integrand is therefore log-concave jointly in $(x_{j+1},x_{j+2:0})$, and Pr\'ekopa's theorem~\cite{Prekopa1973LogConcave} --- if $F:\mathbb{R}^p\times\mathbb{R}^q\to[0,\infty)$ is log-concave then $x_{j+1}\mapsto\int_{\mathbb{R}^q}F(x_{j+1},d_{x'})\,\mathrm{d}d_{x'}$ is log-concave on $\mathbb{R}^p$ --- gives log-concavity of the marginal in $x_{j+1}$.

	\emph{Claim 3. Tilted kernels are $\kappa$-Lipschitz in $W_1$.} Let $y,y'\in\mathcal{Y}$ and let $h:\mathbb{R}^n\to[0,\infty)$ be bounded, log-concave, and not a.e.\ zero on $y'$. For every $x\in y$ and measurable $A\subseteq y'$, define
	\begin{equation}
	\begin{split}
		K(A|x)&:=\frac{\int_A h(x')\,\varphi_n(x';Ax+\hat{w},\Sigma_w)\,\mathrm{d}x'}{\int_{y'}h(x')\,\varphi_n(x';Ax+\hat{w},\Sigma_w)\,\mathrm{d}x'}.
	\end{split}
	\end{equation}
	Then $\int_{y'}h(x')\,\varphi_n(x';Ax+\hat{w},\Sigma_w)\,\mathrm{d}x'\in(0,\infty)$ for every $x\in y$, and, for every $x,\tilde{x}\in y$,
	\begin{equation*}
		W_1\bigl(K(\,\cdot\,|x),K(\,\cdot\,|\tilde{x})\bigr)\le \kappa\,\|x-\tilde{x}\|_2,
	\end{equation*}
	with $\kappa$ as in~\eqref{eq:kappa_definition}. 
	
	\emph{Step 1. Exponential family.} Expanding the Gaussian exponent and absorbing every factor independent of $z$ into the normalization,
	\begin{equation*}
		K(A|x)=\frac{\int_A e^{\theta(x)^\top x'}\eta(\mathrm{d}x')}{\int_{\mathbb{R}^n}e^{\theta(x)^\top \tilde{x}'}\eta(\mathrm{d}\tilde{x}')},
	\end{equation*}
	where $\eta(\mathrm{d}x'):=h(x')\mathbf{1}_{y'}(x')e^{-\frac{1}{2}{x'}^\top\Sigma_w^{-1}x'}\mathrm{d}x'$, and $\theta(x):=\Sigma_w^{-1}(Ax+\hat{w})$.
	As $y'$ is bounded, $h$ is bounded, and $h$ is not a.e.\ zero on $y'$, the denominator is finite and strictly positive for every $\theta$. The map $x\mapsto\theta(x)$ is affine, with linear part $\Sigma_w^{-1}A$.

	\emph{Step 2. Differentiate along the segment.} Fix $x,\tilde{x}\in y$ and interpolate $x_a:=\tilde{x}+a(x-\tilde{x})\in y$ for $a\in[0,1]$, by convexity of $y$. Then $\theta(x_a)$ is affine in $a$ with constant velocity $\dot\theta_a=\Sigma_w^{-1}A(x-\tilde{x})$. Since $x'$ ranges over the bounded set $y'$, differentiation under the integral sign is legitimate and gives, for bounded measurable $f$, $\frac{\mathrm{d}}{\mathrm{d}a}\mathbb{E}_{K(\,\cdot\,|x_a)}[f]=\operatorname{Cov}_{K(\,\cdot\,|x_a)}(f(x'),\dot\theta_a^\top x')$.

	\emph{Step 3. Kantorovich duality and Cauchy--Schwarz.} Both $K(\,\cdot\,|x)$ and $K(\,\cdot\,|\tilde{x})$ are supported on $y'$, so by Kantorovich--Rubinstein duality it suffices to control $\mathbb{E}_{K(\,\cdot\,|x)}[f]-\mathbb{E}_{K(\,\cdot\,|\tilde{x})}[f]$ over bounded $1$-Lipschitz $f:y'\to\mathbb{R}$. By Step 2 and Cauchy--Schwarz,
	\begin{equation*}
		W_1\bigl(K(\,\cdot\,|x),K(\,\cdot\,|\tilde{x})\bigr)\le\sup_{a\in[0,1]}\sup_f\operatorname{Var}_{K(\,\cdot\,|x_a)}(f)^{1/2}\operatorname{Var}_{K(\,\cdot\,|x_a)}(\dot\theta_a^\top x')^{1/2}.
	\end{equation*}

	\emph{Step 4. Two bounds on each variance.} Since 
	\[
	\frac{\mathrm{d}K(\,\cdot\,|x)}{\mathrm{d}\mathcal{N}(\Sigma_w\theta(x),\Sigma_w)}\propto h(x')\mathbf{1}_{y'}(x'),
	\]
	is log-concave, $K(\,\cdot\,|x)$ is a log-concave perturbation of $\mathcal{N}(\Sigma_w\theta(x),\Sigma_w)$, and the Brascamp--Lieb variance inequality~\cite{BrascampLieb1976Extensions} --- for $\gamma=\mathcal{N}(\mu,\Sigma)$ with $\Sigma\succ0$ and $\mathrm{d}\pi\propto e^{-U}\mathrm{d}\gamma$ with $U$ convex, $\operatorname{Var}_\pi(g)\le\mathbb{E}_\pi[\nabla g^\top\Sigma\nabla g]$ for every locally Lipschitz $g\in L^2(\pi)$ --- with $\Sigma=\Sigma_w$ gives, for $1$-Lipschitz $f$, $\operatorname{Var}_{K(\,\cdot\,|x)}(f)\le\|\Sigma_w\|_2$, and $\operatorname{Var}_{K(\,\cdot\,|x)}(\dot\theta_a^\top x')\le\dot\theta_a^\top\Sigma_w\dot\theta_a$. Independently, $K(\,\cdot\,|x)$ is supported on $y'$, so a $1$-Lipschitz $f$ takes values in an interval of length at most $d_M$, and $\dot\theta_a^\top x'$ in one of length at most $\|\dot\theta_a\|_2d_M$; Popoviciu's inequality gives $\operatorname{Var}_{K(\,\cdot\,|x)}(f)\le\frac{1}{4} d_M^2$ and $\operatorname{Var}_{K(\,\cdot\,|x)}(\dot\theta_a^\top x')\le\frac{1}{4}\|\dot\theta_a\|_2^2d_M^2$. Combining the two bounds for each factor, uniformly in $a$,
	\begin{equation*}
	\begin{split}
		\operatorname{Var}_{K(\,\cdot\,|x_a)}(f)^{1/2}&\le\min\{\|\Sigma_w\|_2^{1/2},\frac{1}{2}d_M\},\\
		\operatorname{Var}_{K(\,\cdot\,|x_a)}(\dot\theta_a^\top x')^{1/2}&\le\min\{(\dot\theta_a^\top\Sigma_w\dot\theta_a)^{1/2},\frac{1}{2}\|\dot\theta_a\|_2d_M\}.
	\end{split}
	\end{equation*}
	Substituting $\dot\theta_a=\Sigma_w^{-1}A(x-\tilde{x})$, so that $(\dot\theta_a^\top\Sigma_w\dot\theta_a)^{1/2}=\|\Sigma_w^{-1/2}A(x-\tilde{x})\|_2\le\|\Sigma_w^{-1/2}A\|_2\|x-\tilde{x}\|_2$ and $\|\dot\theta_a\|_2\le\|\Sigma_w^{-1}A\|_2\|x-\tilde{x}\|_2$, and inserting into Step 3 proves the claim, with $\kappa$ as in~\eqref{eq:kappa_definition}.

	\emph{Conclusion.} By Claim 1, $\mu_{\xi,m}^{(\ell)}=\delta_\xi K_{-\ell}^m\cdots K_{-1}^m$, and by Claim 2 each $K_j^m$ is exactly of the form treated in Claim 3, with $y=y_j$, $y'=y_{j+1}$, and $h=b(y_{j+2:0}|\,\cdot\,)$, so, for $x,\tilde{x}\in y_j$,
	\begin{equation*}
	W_1\bigl(K_j^m(\,\cdot\,|x),K_j^m(\,\cdot\,|\tilde{x})\bigr)\le\kappa\|x-\tilde{x}\|_2.
	\end{equation*}
	Such a kernel contracts $W_1$ between arbitrary measures on $y_j$ by the same constant. For $\pi,\widetilde{\pi}\in\mathcal{P}(y_j)$, let $\gamma\in\Gamma(\pi,\widetilde{\pi})$ attain the infimum in the definition~\eqref{eq:wasserstein_distance} of $W_1$ and, for $\gamma$-a.e.\ $(x,\tilde{x})$, let $\gamma_{x,\tilde{x}}\in\Gamma\bigl(K_j^m(\,\cdot\,|x),K_j^m(\,\cdot\,|\tilde{x})\bigr)$ attain it likewise, chosen measurably. The mixture $\int\gamma_{x,\tilde{x}}\,\gamma(\mathrm{d}x,\mathrm{d}\tilde{x})$ of the conditional couplings $\gamma_{x,\tilde{x}}$ against $\gamma$ is a coupling of $\pi K_j^m$ and $\widetilde{\pi}K_j^m$ with cost
	\begin{equation*}
		\int W_1\bigl(K_j^m(\,\cdot\,|x),K_j^m(\,\cdot\,|\tilde{x})\bigr)\,\gamma(\mathrm{d}x,\mathrm{d}\tilde{x})\le\kappa\int\|x-\tilde{x}\|_2\gamma(\mathrm{d}x,\mathrm{d}\tilde{x})=\kappa\,W_1(\pi,\widetilde{\pi}),
	\end{equation*}
	so $W_1(\pi K_j^m,\widetilde{\pi}K_j^m)\le\kappa\,W_1(\pi,\widetilde{\pi})$. Applying this $\ell$ times, starting from $W_1(\delta_\xi,\delta_\zeta)=\|\xi-\zeta\|_2\le d_M$ for admissible $\xi,\zeta\in y_{-\ell}$, gives $\Delta_\ell(m)\le\kappa^\ell d_M$. Since $\mu_{\xi,m}^{(\ell)}$ and $\mu_{\zeta,m}^{(\ell)}$ are both supported on the common cell $y_0$ of diameter $d_M$, any coupling also gives $\Delta_\ell(m)\le d_M$ trivially. Together these prove~\eqref{eq:gaussian_dispersion_bound}.

	\emph{Isotropic noise.} For $\Sigma_w=\sigma^2I_n$, the first factor in~\eqref{eq:kappa_definition} is $\min\{\sigma,d_M/2\}$ and the second is $\|A\|_2\min\{\sigma^{-1},d_M/(2\sigma^2)\}$. If $d_M\ge2\sigma$, both minima are attained by $\sigma$ and $\sigma^{-1}$, giving $\|A\|_2$; if $d_M<2\sigma$, they are attained by $d_M/2$ and $d_M/(2\sigma^2)$, giving $\|A\|_2d_M^2/(4\sigma^2)$.

	Finally, applying Theorem~\ref{thm:absorbing_value_function_width_convergence} with~\eqref{eq:gaussian_density_sensitivity_bound} and~\eqref{eq:gaussian_dispersion_bound} gives~\eqref{eq:gaussian_value_function_width_bound}.
\end{pf*}

% \begin{pf*}{Proof of Proposition~\ref{prop:kappa_sharp}}
% 	Here $\mu_{\xi,m}^{(1)}$ is $\mathcal{N}(a\xi,\sigma^2)$ conditioned on $[0,d_M]$. For $\xi\in(0,d_M)$ we have $a\xi\in(0,ad_M)\subset(0,d_M)$, an interior point, so $\mu_{\xi,m}^{(1)}\Rightarrow\delta_{a\xi}$ as $\sigma\downarrow0$; since all measures are supported on the bounded set $[0,d_M]$, weak convergence implies $W_1$-convergence. Hence $\Delta_1\to\sup_{\xi,\zeta\in[0,d_M]}|a\xi-a\zeta|=ad_M$. As $\sigma\downarrow0$ we also have $d_M^2/(4\sigma^2)\to\infty$, so $\kappa\to a$ by Proposition~\ref{prop:absorbing_linear_gaussian}.
% \qed
% \end{pf*}

\bibliographystyle{plain}
\bibliography{myreferences}

\end{document}